\documentclass[aip,reprint,apl, amsmath,amssymb]{revtex4-1}

\usepackage{graphicx}
\usepackage{dcolumn}
\usepackage{bm}
\usepackage[colorlinks=true,citecolor=blue,urlcolor=blue,linkcolor=black]{hyperref} 

\renewcommand{\url}[1]{\href{#1}{Link}}

\usepackage[caption=false]{subfig}
\usepackage[acronym,shortcuts,acronymlists={hidden}]{glossaries}
\usepackage{xcolor}
\usepackage{xspace}

\newcommand{\tophat}{top-hat\xspace}
\newcommand{\beamshaper}{beamshaper\xspace}
\newcommand{\asphericon}{Asphericon\xspace}

\newacronym{pv}{PV}{peak-valley}
\newacronym{lmt}{LMT}{Large Momentum Transfer}

\begin{document}


\title{Atom Interferometry with Top-Hat Laser Beams}

\author{N. Mielec}
\author{M. Altorio} 
\author{R. Sapam}
\affiliation{LNE-SYRTE, Observatoire de Paris, Universit\'e PSL, CNRS, Sorbonne Universit\'e, 61 Avenue de l'Observatoire, 75014 Paris, France}
\author{D. Horville}
\affiliation{GEPI, Observatoire de Paris, Universit\'e PSL, CNRS,  5 Place Jules Janssen, 92190 Meudon, France}
\author{D. Holleville}
\author{L.A. Sidorenkov}
\author{A. Landragin}
\author{R. Geiger}
\email{remi.geiger@obspm.fr}
\affiliation{LNE-SYRTE, Observatoire de Paris, Universit\'e PSL, CNRS, Sorbonne Universit\'e, 61 Avenue de l'Observatoire, 75014 Paris, France}

\date{\today}

\begin{abstract}
The uniformity of the intensity and phase of laser beams is crucial to  high-performance atom interferometers. Inhomogeneities in the laser intensity profile cause contrast reductions and systematic effects in interferometers operated with atom sources at micro-Kelvin temperatures, and detrimental diffraction phase shifts in interferometers using large momentum transfer beam splitters. We report on the implementation of a so-called \tophat laser beam in a long-interrogation-time cold-atom interferometer to overcome the issue  of the inhomogeneous laser intensity encountered when using Gaussian laser beams.  We characterize the intensity and relative phase profiles of the \tophat beam and demonstrate  its gain in atom-optics efficiency  over a Gaussian beam, in agreement with numerical simulations. We discuss the application of \tophat beams to improve the performance of different architectures of atom interferometers. 

\end{abstract}

\maketitle




\label{par:intro}
Inertial sensors based on light-pulse atom interferometry address various applications ranging from inertial navigation \cite{Canuel2006,Geiger2011,Cheiney2018}, metrology \cite{Bouchendira2011,Rosi2014,Parker2018}, gravimetry \cite{Peters2001,Hu2013,Gillot2014,Freier2016,Wang2018,Bidel2018,Karcher2018} and gradiometry \cite{McGuirk2002,Sorrentino2012}, tests of fundamental physics \cite{Dimopoulos2007,Lepoutre2012,Aguilera2014,Zhou2015,Jaffe2017,Asenbaum2017}, or gravitational wave astronomy \cite{Chaibi2016,Hogan2016}. 
Light-pulse atom interferometers rely on the coherent transfer of momentum from the photons of counter-propagating laser beams to free falling atoms in order to split, deflect and recombine the matter-waves. 
The sensitivity and accuracy of the instruments thus crucially depend on the relative phase uniformity of the laser beams realizing these atom-optics functionalities. 
State-of-the-art  cold-atom sensors  typically use sources at few $\mu$K temperatures, interrogation times of  several hundreds of milliseconds, and two-photon transitions \cite{Rosi2014,Freier2016,Dutta2016}.
Inhomogeneities in the laser intensity across the atom  cloud degrade the atom optics efficiency, which causes a decrease of interferometer contrast and hence a lower signal to noise ratio, as well as systematic effects \cite{Gauguet2009}. 
Such detrimental effects are amplified in interferometers employing large momentum transfer (LMT) techniques (in which several momenta are transferred to the atoms)  \cite{Asenbaum2017,Mazzoni2015}, in particular because of diffraction phase shifts  \cite{Buechner2003}.
The problem of intensity inhomogeneity can be mitigated by employing  Gaussian beams with a size much larger than that of the atom cloud, at the cost of a reduced peak intensity.

In this work, we report on the implementation of a \textit{collimated}  \tophat laser beam (i.e. with a uniform intensity distribution in the central part \cite{Gori1994}) as a solution to circumvent the problems encountered in atom interferometers employing Gaussian beams.

Beamshaping is a topic of intense development, with applications ranging from micro-lithography, optical data storage, or optical tweezers, where  different approaches are followed to produce structured light patterns. 
For application to atom interferometry, the requirement on the relative phase homogeneity motivates a scheme where  the counter-propagating beam pair is obtained by retro-reflection (the retro-distance typically lying in the ten-centimeters-to-meter scale).
The  interrogation laser beams are thus required to be well collimated over such distances. This requirement on the beam shaping technique amounts to achieving  a flat phase profile.

The simplest form of shaping the intensity distribution of a laser beam, apodization,  results in significant loss of optical power (for example, the optimal transformation of a Gaussian beam into a beam with a  flat intensity profile sacrifies  $64\%$ of the  power).
More efficient techniques involve diffractive optical elements, such as spatial light modulators (SLMs),  in order to produce focused light patterns \cite{Pal2018}, or collimated structured beams when multiple SLMs are cascaded \cite{Ma2010}.
However,  the bulkiness of the optical setup, the potential drift of the beam-shaping performance linked to the use of an active material, and the limited incident peak intensity  make such solutions cumbersome for  atom interferometry experiment.
Instead, passive refractive techniques based on aspheric optical elements \cite{Hoffnagle2000}  seem   favorable,  owing to their compactness, stability, and efficiency.


\begin{figure*}[!ht]
	\centering
	\includegraphics[width=\linewidth]{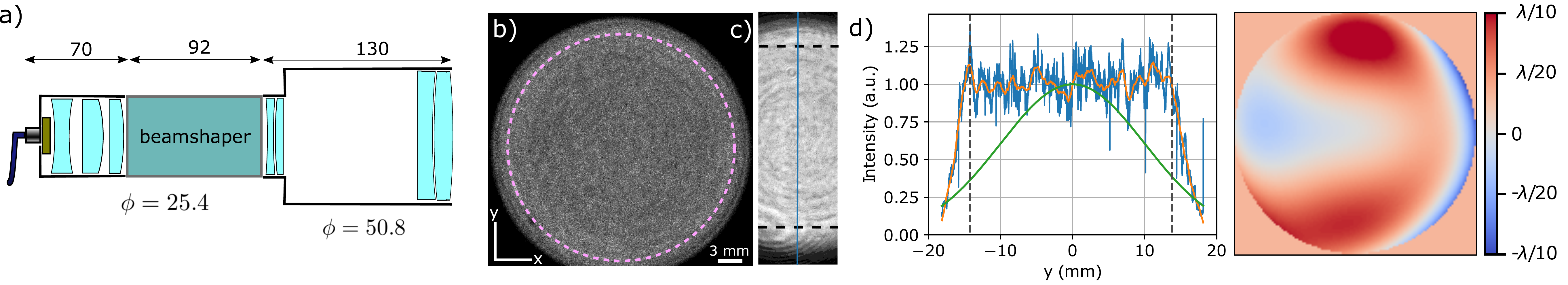}
	\caption{
	a) Schematic view of the optical system with the input  collimator, the beamshaper, and the expander  (dimensions in mm, $\phi$ denoting the diameter of the optics). 
	b) Image of the \tophat beam  on a paper screen. The dashed purple line is a circle of 28~mm diameter. 
	c) Image obtained with a beamprofiler, after 40~cm of propagation.	Between the 2 dashed lines separated by 28~mm,	the uniformity of the plateau  is 0.11 rms.
	d) (blue) Vertical line profile of the \tophat beam shown in c); the intensity has been normalized to the mean plateau intensity. (orange) Moving average over 1~mm.
		(green) Theoretical profile of a Gaussian beam with 40~mm $1/e^2$ diameter.
	e) Relative phase of the \tophat beam with 70~cm propagation difference, in a disk of 28~mm; the  deviation is $\lambda / 5$ (PV) and  $\lambda/28$ (rms). 
	}
	\label{fig:tophat_charact_full}
\end{figure*}

\label{par:optical_charact}
Our passive \tophat collimator solution is based on a recently released commercial  \beamshaper from the \asphericon company (model TSM-25-10-S-B), see Fig.~\ref{fig:tophat_charact_full}a).
The \beamshaper shall receive at its input a Gaussian beam of 10~mm $1/e^2$-diameter    and produce a \tophat beam of 15~mm full width at half maximum (FWHM), with a region of about 14~mm where the intensity varies by less than $10\%$ (Ref.~\cite{asphericonWebsite}).
The beamshaping is done with multiple aspheric optics, based on principles similar to those of Ref.~\cite{Hoffnagle2000}.
The advertized  uniformity of intensity plateau is 0.056 rms, with a phase inhomogeneity of  $\lambda/3$ \ac{pv} and $\sim \lambda / 20$ rms, allowing the beam to propagate without deformation on distances of several meters \cite{asphericonWebsite}.
We inject the \beamshaper with a home-made fiber collimator made of 3 simple lenses, to produce a Gaussian beam of $9.95\pm 0.05$~mm $1/e^2$ diameter. At the output of the \beamshaper, the \tophat beam is magnified by a factor of two with two achromatic doublets, in order to reach a useful region of 28~mm. 
The optical system  can be mounted conveniently on an experiment. 
The power transmission of the input collimator  plus the beamshaper is $91\%$, while that of the full system is $85\%$.
The quality of the generated \tophat beam  mainly depends on the input beam size (which must fall within the 10~mm diameter specification at the $10\%$ level \cite{asphericonWebsite}), and of its collimation.

To align the \tophat collimator, we image the beam on a paper screen, and optimize the intensity profile by moving the input fiber placed on a 5-axis mount. 
We target a flat circular intensity profile maintained over a propagation distance of  at least 150~cm.
Fig~\ref{fig:tophat_charact_full}b) shows the  beam imaged on the paper screen at the output of the expander. While this method is convenient for the alignment procedure, it is not suited for a precise measurement of the intensity uniformity of the beam because of the speckle produced on the paper screen.
We use a large-area beamprofiler ($11.3\times 6.0$~mm$^2$) to measure  the uniformity of the plateau.
Fig~\ref{fig:tophat_charact_full}c) shows the stitched images acquired by scanning the beamprofiler in front of the beam after 40~cm of propagation.
The beam exhibits a qualitatively flat plateau.
Large diameter rings concentric to the beam  are attributed to the \beamshaper.
The uniformity of the plateau over a diameter of 28~mm is 0.11 rms, and the FWHM is $31.7\pm0.2$~mm.
Fig~\ref{fig:tophat_charact_full}d) shows a  profile of the vertical cut  through the middle of the beam (along the blue line).
The orange line is a moving average over 1~mm of the profile, shown here to illustrate lower frequency inhomogeneities.
For comparison, the green line shows a Gaussian beam with 40~mm diameter at $1/e^2$ (as used in Ref.~\cite{Dutta2016}) and same peak intensity as the \tophat beam.

In an atom interferometer, the relative phase between  two counter-propagating laser beams  is imprinted on the atomic wave-function during the light pulses. This relative phase contains a term associated with the free propagation, $\varphi(x,y,0)-\varphi(x,y,2L)$, where  $L$ the  distance between the atom cloud and the retro-mirror \cite{Louchet-Chauvet2011}. 
We measured such relative phase field for our \tophat beam using an asymmetric Michelson  interferometer  with the difference of its arms set to $2L$.
At the output, the interference pattern carries the 2D relative phase map, which  we recover  using a Fourier  analysis \cite{Takeda1982}.
A lower bound on the accuracy is set by the planeity of the mirrors and of the beamsplitter used in the interferometer, specified to be $\lambda/10$ peak-valley (PV).
The relative phase map  in a pupil of 28~mm diameter corresponding to the useful part of the beam is shown Fig~\ref{fig:tophat_charact_full}e), for a difference in propagation distance $2L=70$~cm. We find relative phase  inhomogeneities of $\lambda / 5$ PV and a $\lambda / 28$ rms.
Additional phase maps for further propagation distances are given in the supplemental material.
Our characterizations  show that the \tophat beam is suitable for high-precision atom interferometry, where  relative wavefront inhomogeneities are an issue \cite{Gauguet2009,Louchet-Chauvet2011,Schkolnik2015,Karcher2018}.

\begin{figure}[!h]
	\centering
	\includegraphics[width=\linewidth]{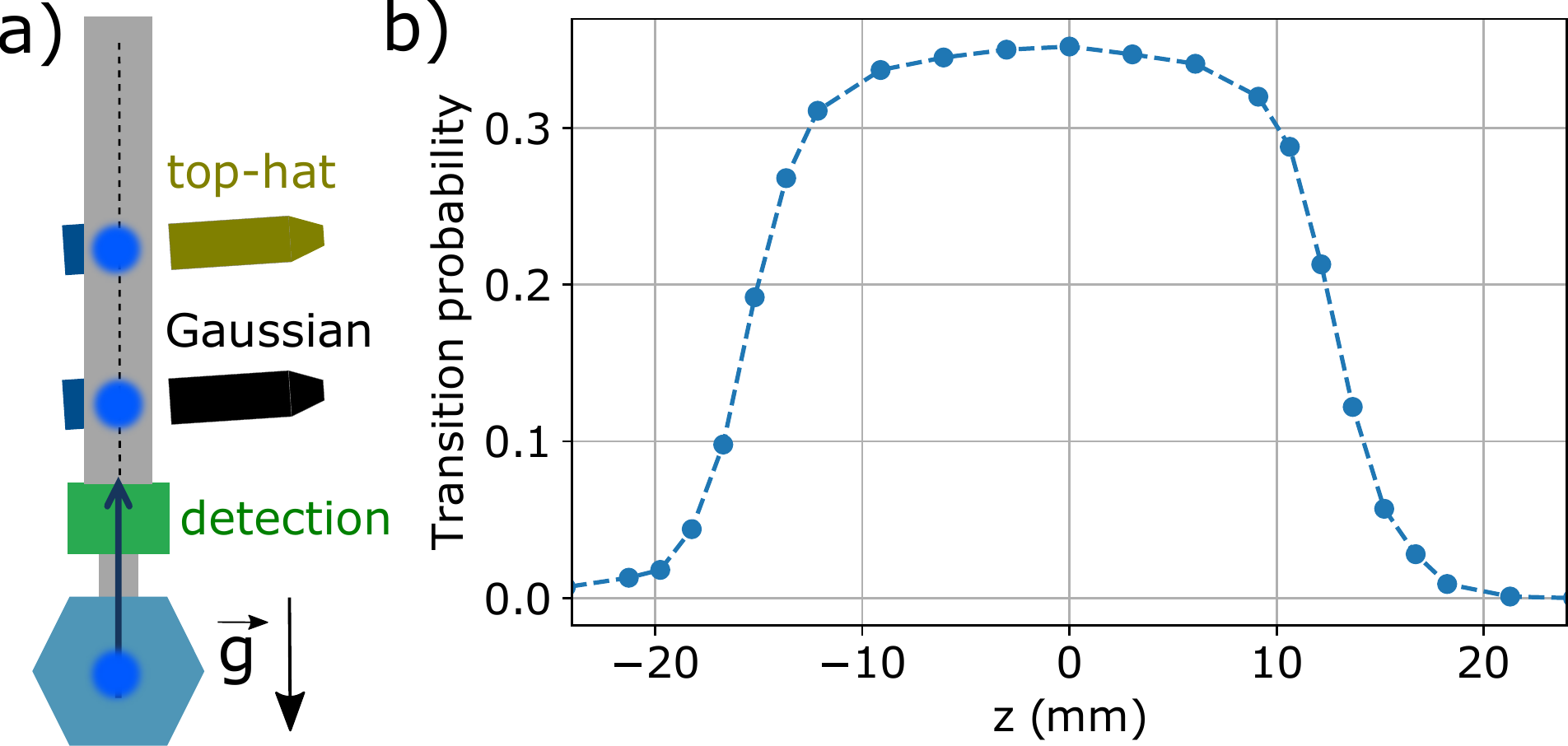}
	\caption{a) Sketch of the experiment.
	b)  Measurement of the local Raman lasers intensity with a cold atom cloud, by recording the transition probability versus time-of-flight. 	
	The duration of the Raman pulse is fixed ($\tau=9 \ \mu$s) and set close to that of a  $\pi/2$ pulse, where the sensitivity to intensity fluctuations on the plateau is maximum.
	 The horizontal axis ($z$) is obtained by multiplying the TOF with the mean velocity of the atoms in the beam (3.0~m.s$^{-1}$).  
	}
	\label{fig:upwards_atoms}
\end{figure}

We implemented the \tophat beam on a cold-atom gyroscope-accelerometer experiment. The setup has been described in previous works \cite{Meunier2014,Dutta2016} and we recall here the main features which are relevant for this study.
Laser-cooled Cesium atoms (temperature of $1.2 \ \mu$K) are launched  vertically with a  velocity of up to $5.0$~m.s$^{-1}$. After a  selection  step of the $m_F=0$ magnetic sublevel, we realize the atom interferometer by means of two-photon stimulated Raman transitions from counter-propagating laser beams, which couple the $|F=3,m_F=0\rangle$ and $|F=4,m_F=0\rangle$ clock states.
 The direction of the Raman beams is nearly horizontal. We use two beams separated vertically by a distance of 211~mm. The \tophat collimator was set up at the position of the top beam, while the bottom beam is a Gaussian beam of 40~mm diameter at $1/e^2$ (Fig.~\ref{fig:upwards_atoms}a)).
The   state of the atoms at the output of the interferometer is finally read out  using fluorescence detection.

We first probe the intensity profile of the \tophat beam by applying a Raman pulse of fixed duration $\tau$ at different times as the atoms travel on their way up. The atoms are launched with velocity of  4.7~m.s$^{-1}$, and their mean trajectory intersects the center of the beam after a time of flight (TOF) of 170~ms. After this relatively short TOF, the  size of the cloud is still close to that of the initially launched atoms ($\simeq 1.5$~mm rms radius) and much  smaller than the beam size. The transition probability, $P\propto \sin^2(\Omega(z)\tau/2)$, is determined by the local value of the two-photon Rabi frequency, $\Omega(z)$, and can thus be used as a probe of the local intensity of the beam (here $z$ denotes the direction parallel to gravity). 
Fig.~\ref{fig:upwards_atoms}b) shows the transition probability versus the relative position of the cloud inside the beam. We observe a qualitatively  flat intensity profile in the center, with a width  consistent with the optical characterization reported in Fig.~\ref{fig:tophat_charact_full}. 

\begin{figure}[!h]
	\centering
	\includegraphics[width=0.9\linewidth]{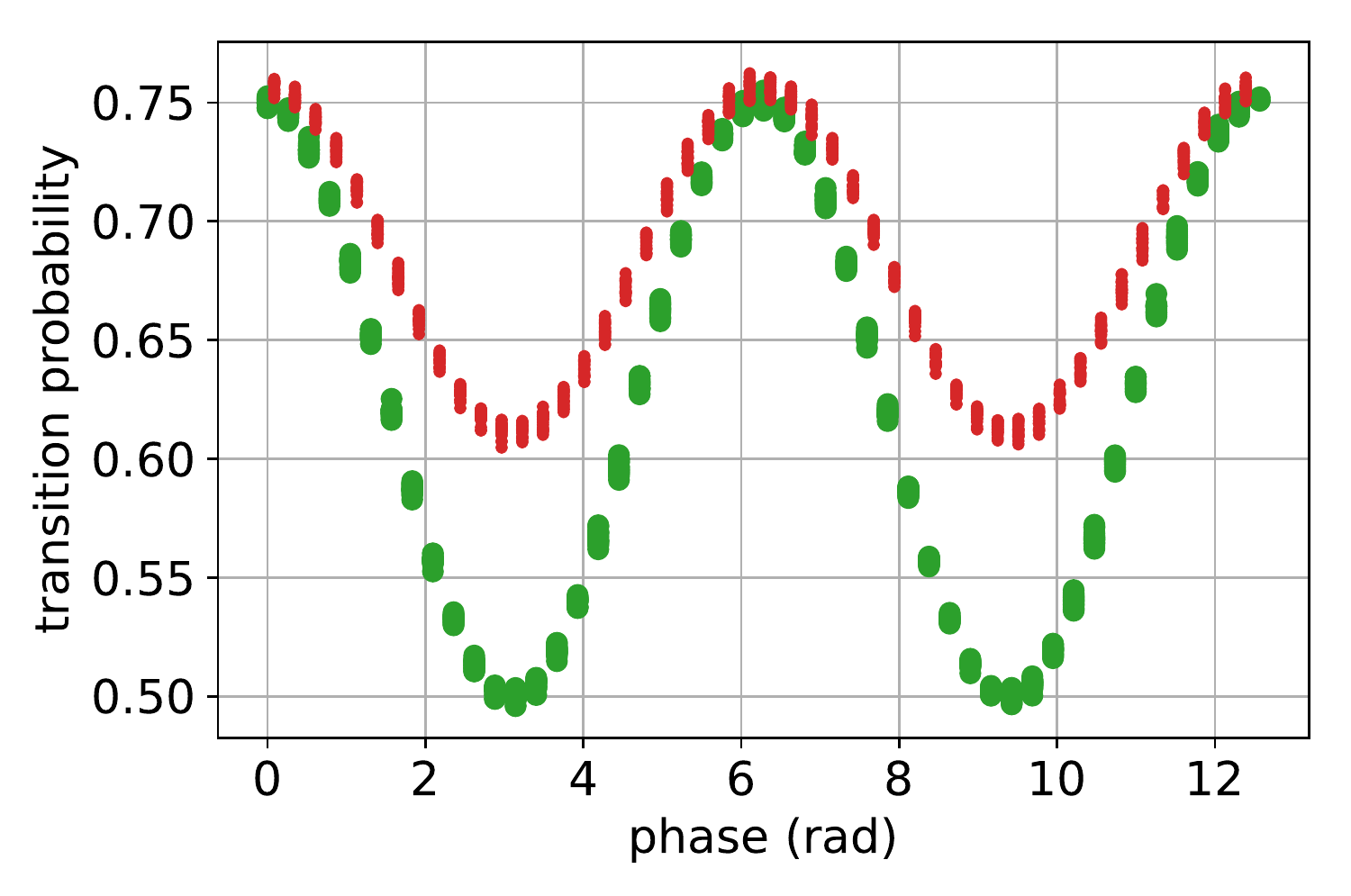}
	\caption{Interference fringes for a 3-pulse interferometer sequence with a pulse separation time  $T=1$~ms, after a TOF of 855~ms.  Red: Gaussian beam. Green: \tophat beam.
	The interference fringes are scanned by varying the relative Raman laser phase at the third light pulse.
	The same optical power was used for the Gaussian and the \tophat beams.
	}
	\label{fig:fringes_short_T}
\end{figure}

The size of a cold atom cloud increases over free propagation due to finite temperature. This results in an inhomogeneous atom-light coupling when the cloud size approaches the waist of the Gaussian beam, thereby decreasing the interferometer contrast. 
The intensity homogeneity of the \tophat beam allows in principle to improve on this effect. To illustrate this improvement, we operate a 3 light-pulse interferometer sequence  with a pulse separation time $T=1$~ms, after a long TOF of 855~ms to bring forward the effect of the atom cloud expansion.
For a quantitative comparison, the difference in height between the two beams (211~mm) was matched  by the respective change in launch velocity, in order to obtain  nearly the same TOFs when crossing the Gaussian and \tophat beams.
  Fig.~\ref{fig:fringes_short_T} presents the comparison and shows the advantage of the \tophat beam.

\begin{figure}
	\centering
	\includegraphics[width=\linewidth]{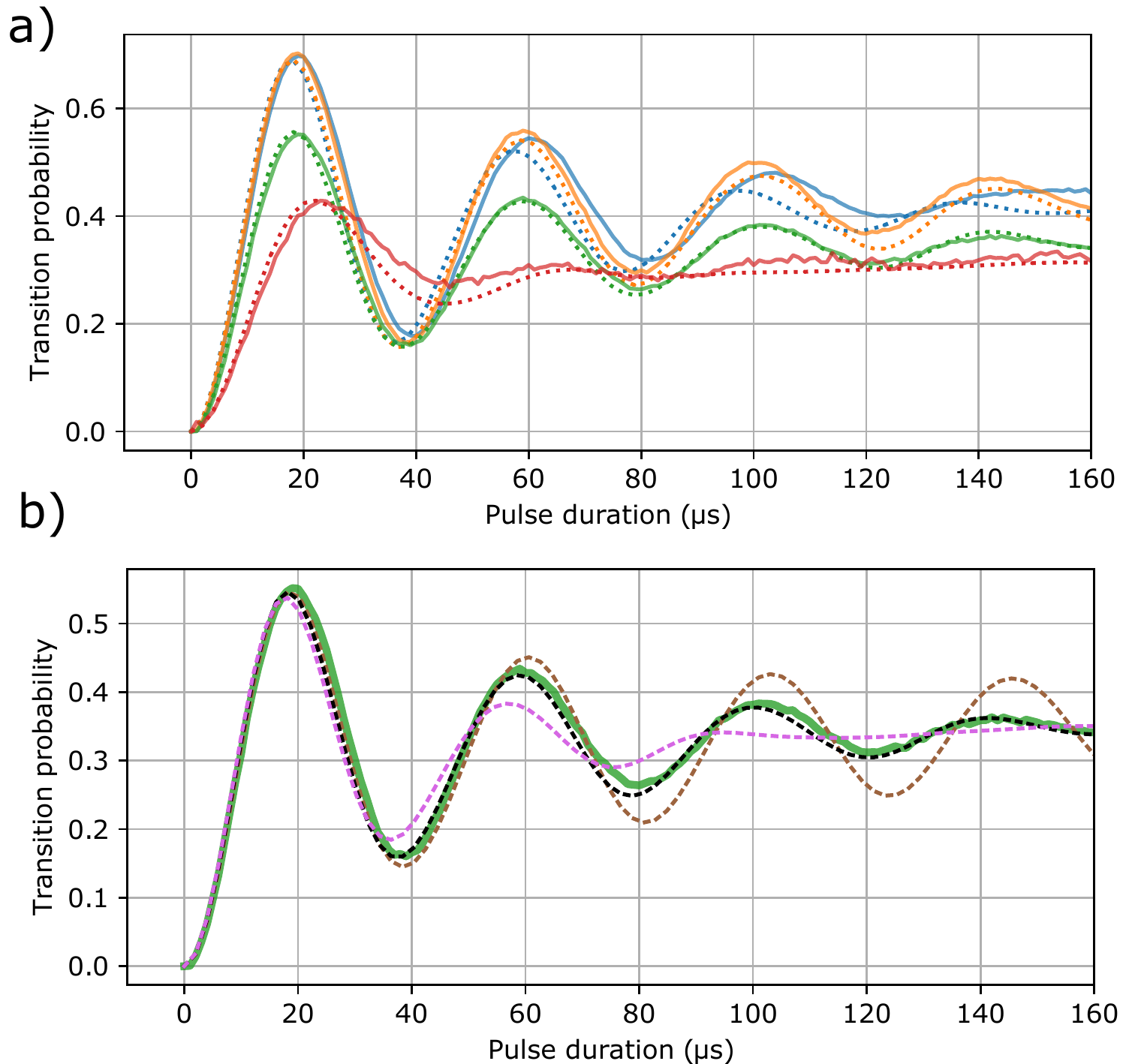}
	\caption{Rabi oscillations. a) Plain lines: measured oscillations on the way up after 170~ms of TOF (blue and orange for Gaussian and \tophat, respectively), and on the way down after 855~ms of TOF (red and green). Dotted lines: numerical simulation. 
	b) Green plain line: measured Rabi oscillation in the \tophat beam after 855~ms of TOF (the same as in a)). Dashed: numerical simulation for various level of rms intensity noise on the \tophat (brown: $0\%$, black: $8.3\%$, violet: $15\%$). 
	}
	\label{fig:rabi_oscillations}
\end{figure}

To assess  the limitations to the gain in atom-optics efficiency  offered by our \tophat beam over our Gaussian beam, we recorded Rabi oscillations after various TOF, when the launched atom cloud crosses the beams on its  way up and on its way down. 
Fig~\ref{fig:rabi_oscillations}a) shows the Rabi oscillations on the way up after a TOF of 170~ms and on the way down after TOF of 855~ms for the \tophat and Gaussian beams. On the way up, the cloud size is smaller than the beam sizes, and the Rabi oscillations have a similar shape for the Gaussian and \tophat beams, as expected. The  transfer efficiency of $\sim 70\%$ is limited  by the velocity selectivity of the two-photon transition, given by the finite Rabi frequency (i.e. laser power) and velocity spread of the atoms in the direction of the beams.
 On the contrary, on the way down, the Rabi oscillation in the \tophat beam (green) is significantly improved with respect to that in the Gaussian beam (red), owing to the homogeneity of the two-photon Rabi frequency from the \tophat beam. 
To model the Rabi oscillations, we employ a  Monte-Carlo simulation where we generate an ensemble of atoms with individual velocities following the   distribution measured with the Doppler-sensitive Raman transitions (corresponding to a 3D temperature of $1.2 \ \mu$K), and propagate them in the Raman beams. 
The details of the model are given in the Supplementary   Material.
The model reproduces well the  data, and allows to assess the residual intensity inhomogeneities of the \tophat beam.
Fig.~\ref{fig:rabi_oscillations}b) shows the measured Rabi oscillation confronted to a simulation where intensity noise of various levels is added on the \tophat profile \cite{noteSpatialFreq}. The  data match best  the numerical simulation assuming an inhomogeneity of $8.3\%$ rms, consistent with the optical characterization of the intensity inhomogeneities of $11\%$ reported in Fig.~\ref{fig:tophat_charact_full}.

\begin{figure}[!h]
	\centering
	\includegraphics[width=\linewidth]{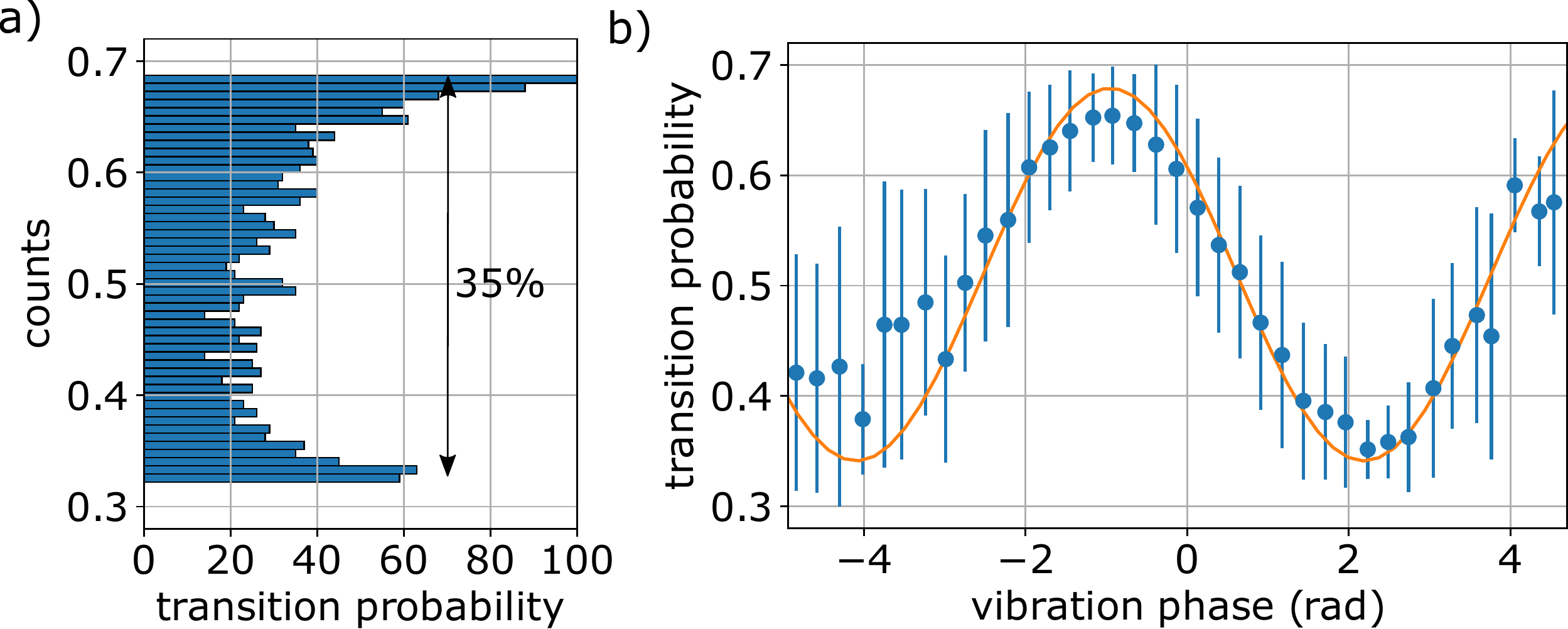}
	\caption{Performance of the 3-pulse interferometer with  $2T=294$~ms.  
	a) Histogram of the transition probability showing a contrast of $35\%$. 
	b) Transition probability versus the vibration phase calculated from the data of a broadband seismometer. The  data points are sorted along the  x-axis and binned in intervals of 262~mrad width. 
	The imperfect estimation of the vibration phase by the seismometer translates into phase noise and consequently into probability noise when binning. Error bars denote 1 standard deviation.
	The orange plain curve shows a sinusoidal fit where the fringe amplitude is set to the value of $35\%$ extracted from a).
		}
	\label{fig:fringes_long_T}
\end{figure}

Finally, we demonstrate that the \tophat beam is suited for high-sensitivity atom interferometry, by running a 3-pulse atom interferometer sequence with a pulse separation time $T=147$~ms. The first $\pi/2$ pulse is realized in the Gaussian beam (on the way up, $\text{TOF}=170$~ms), while the second and third pulses are realized in the \tophat beam ($\text{TOF}=317$ and $464$~ms). 
For such long interrogation time, the interferometer is highly sensitive to vibration noise producing at its output a typical rms phase shift of more than $\pi$ rad. Running the interferometer results in a random sampling of the fringe pattern by vibration noise, which appears blurred  without additional knowledge on  vibration noise at each run. To extract the contrast, we follow the method of Ref.~\cite{Geiger2011} and compute the histogram of the transition probability data (Fig.~\ref{fig:fringes_long_T}a), from which we extract a contrast of $35\%$.
Furthermore, we recover the interference fringes by correlating the atom interferometer output with the phase calculated from vibration data acquired with two brodband seismometers \cite{Merlet2009,Dutta2016}, see Fig.~\ref{fig:fringes_long_T}b). 
The uncertainty ($1\sigma$) on the fitted phase is 80~mrad, corresponding to an horizontal acceleration uncertainty of $2.5\times 10^{-7}$~m.s$^{-2}$. Although the measurement sensitivity is limited by residual vibration noise, this experiment shows that the \tophat beam is compatible with high-sensitivity inertial measurements based on long-interrogation-time cold-atom interferometry.

\label{par:discussion}
In conclusion, we have set up and characterized a collimated \tophat laser beam and reported on its implementation in a long interrogation time cold-atom interferometer. Our \tophat beam  features a constant intensity over a region of $28$~mm with rms variations of about $10\%$,  
We expect that the intensity homogeneity offered by \tophat beams compared to Gaussian beams will be beneficial to various atom interferometer geometries which we discuss below. We present additional advantages  in the Supplementary Material.

The intensity homogeneity of the interrogation beams will  allow reducing or canceling important systematic effects in cold-atom interferometers,  such as the  two photon light shift \cite{Gauguet2008}. It can also be used to improve the efficiency and stability of atom launching techniques based on the coherent transfer of photon momenta, such as in Bloch oscillations \cite{Bouchendira2011,Parker2018,Asenbaum2017}.
 Moreover, this beamshaping solution could be adapted for atom interferometers with baselines of several meters as in Ref.~\cite{Asenbaum2017}.

Employing a single \tophat beam can  be used to build compact, yet precise, cold-atom inertial sensors. For example, a $D=28$~mm wide homogeneous intensity profile should allow to run a fountain interferometer  with a total interferometer time $2T\simeq 2\times\sqrt{2D/g}= 151$~ms if the atoms are launched from the bottom  of the beam. 
Moreover, the design of gyroscopes, where the atoms travel through successive laser beams with a velocity transverse to the momentum transfers \cite{Gauguet2009,Tackmann2012,Yao2018}, could  be simplified with a single \tophat beam.
  

Homogeneity of the intensity profile should  reduce the diffraction phase shifts encountered in LMT Bragg diffraction \cite{Chiow2011,Altin2013,Estey2015}. For example, a variation of $1\%$ of laser intensity in $4\hbar k$ Bragg diffraction  amounts to a variation in diffraction phase  of about 84~mrad \cite{Buechner2003}.
The rms intensity uniformity of our \tophat beam is between $8\%$ and $11\%$ over a region of $28$~mm (Fig.~\ref{fig:tophat_charact_full}c). 
Keeping a $10\%$ rms intensity variation within a Gaussian beam  requires to work within a reduced portion  around the center, which translates in using only $25\%$ of the total power. This suggests that the efficiency and accuracy of LMT beam splitters should be significantly improved by employing \tophat beams.


\begin{acknowledgments}
\label{par:Acknwoledgments}
We thank Josiane Firminy and Faouzi Boussaha for their realization of engraved aspheric phase plates in an early design of beamshaper conducted in the beginning of this project. This work was supported by Ville de Paris (project HSENS-MWGRAV), FIRST-TF (ANR-10-LABX-48-01),   Centre National d'Etudes Saptiales (CNES), Sorbonne Universit\'es (project LORINVACC), Action Sp\'ecifique du CNRS Gravitation, R\'ef\'erences, Astronomie et M\'etrologie (GRAM), and by the European Union's Horizon 2020 research and innovation programme under the Marie Sklodowska-Curie grant agreement No 660081. We thank Laurence Pruvost for fruitful discussions.
\end{acknowledgments}


\bibliography{biblio_tophat}

\begin{thebibliography}{45}%
\makeatletter
\providecommand \@ifxundefined [1]{%
 \@ifx{#1\undefined}
}%
\providecommand \@ifnum [1]{%
 \ifnum #1\expandafter \@firstoftwo
 \else \expandafter \@secondoftwo
 \fi
}%
\providecommand \@ifx [1]{%
 \ifx #1\expandafter \@firstoftwo
 \else \expandafter \@secondoftwo
 \fi
}%
\providecommand \natexlab [1]{#1}%
\providecommand \enquote  [1]{``#1''}%
\providecommand \bibnamefont  [1]{#1}%
\providecommand \bibfnamefont [1]{#1}%
\providecommand \citenamefont [1]{#1}%
\providecommand \href@noop [0]{\@secondoftwo}%
\providecommand \href [0]{\begingroup \@sanitize@url \@href}%
\providecommand \@href[1]{\@@startlink{#1}\@@href}%
\providecommand \@@href[1]{\endgroup#1\@@endlink}%
\providecommand \@sanitize@url [0]{\catcode `\\12\catcode `\$12\catcode
  `\&12\catcode `\#12\catcode `\^12\catcode `\_12\catcode `\%12\relax}%
\providecommand \@@startlink[1]{}%
\providecommand \@@endlink[0]{}%
\providecommand \url  [0]{\begingroup\@sanitize@url \@url }%
\providecommand \@url [1]{\endgroup\@href {#1}{\urlprefix }}%
\providecommand \urlprefix  [0]{URL }%
\providecommand \Eprint [0]{\href }%
\providecommand \doibase [0]{http://dx.doi.org/}%
\providecommand \selectlanguage [0]{\@gobble}%
\providecommand \bibinfo  [0]{\@secondoftwo}%
\providecommand \bibfield  [0]{\@secondoftwo}%
\providecommand \translation [1]{[#1]}%
\providecommand \BibitemOpen [0]{}%
\providecommand \bibitemStop [0]{}%
\providecommand \bibitemNoStop [0]{.\EOS\space}%
\providecommand \EOS [0]{\spacefactor3000\relax}%
\providecommand \BibitemShut  [1]{\csname bibitem#1\endcsname}%
\let\auto@bib@innerbib\@empty
\bibitem [{\citenamefont {Canuel}\ \emph {et~al.}(2006)\citenamefont {Canuel},
  \citenamefont {Leduc}, \citenamefont {Holleville}, \citenamefont {Gauguet},
  \citenamefont {Fils}, \citenamefont {Virdis}, \citenamefont {Clairon},
  \citenamefont {Dimarcq}, \citenamefont {Bord\'e}, \citenamefont {Landragin},\
  and\ \citenamefont {Bouyer}}]{Canuel2006}%
  \BibitemOpen
  \bibfield  {author} {\bibinfo {author} {\bibfnamefont {B.}~\bibnamefont
  {Canuel}}, \bibinfo {author} {\bibfnamefont {F.}~\bibnamefont {Leduc}},
  \bibinfo {author} {\bibfnamefont {D.}~\bibnamefont {Holleville}}, \bibinfo
  {author} {\bibfnamefont {A.}~\bibnamefont {Gauguet}}, \bibinfo {author}
  {\bibfnamefont {J.}~\bibnamefont {Fils}}, \bibinfo {author} {\bibfnamefont
  {A.}~\bibnamefont {Virdis}}, \bibinfo {author} {\bibfnamefont
  {A.}~\bibnamefont {Clairon}}, \bibinfo {author} {\bibfnamefont
  {N.}~\bibnamefont {Dimarcq}}, \bibinfo {author} {\bibfnamefont {C.~J.}\
  \bibnamefont {Bord\'e}}, \bibinfo {author} {\bibfnamefont {A.}~\bibnamefont
  {Landragin}}, \ and\ \bibinfo {author} {\bibfnamefont {P.}~\bibnamefont
  {Bouyer}},\ }\href {\doibase 10.1103/PhysRevLett.97.010402} {\bibfield
  {journal} {\bibinfo  {journal} {Phys. Rev. Lett.}\ }\textbf {\bibinfo
  {volume} {97}},\ \bibinfo {pages} {010402} (\bibinfo {year}
  {2006})}\BibitemShut {NoStop}%
\bibitem [{\citenamefont {Geiger}\ \emph {et~al.}(2011)\citenamefont {Geiger},
  \citenamefont {M\'enoret}, \citenamefont {Stern}, \citenamefont {Zahzam},
  \citenamefont {Cheinet}, \citenamefont {Battelier}, \citenamefont {Villing},
  \citenamefont {Moron}, \citenamefont {Lours}, \citenamefont {Bidel},
  \citenamefont {Bresson}, \citenamefont {Landragin},\ and\ \citenamefont
  {Bouyer}}]{Geiger2011}%
  \BibitemOpen
  \bibfield  {author} {\bibinfo {author} {\bibfnamefont {R.}~\bibnamefont
  {Geiger}}, \bibinfo {author} {\bibfnamefont {V.}~\bibnamefont {M\'enoret}},
  \bibinfo {author} {\bibfnamefont {G.}~\bibnamefont {Stern}}, \bibinfo
  {author} {\bibfnamefont {N.}~\bibnamefont {Zahzam}}, \bibinfo {author}
  {\bibfnamefont {P.}~\bibnamefont {Cheinet}}, \bibinfo {author} {\bibfnamefont
  {B.}~\bibnamefont {Battelier}}, \bibinfo {author} {\bibfnamefont
  {A.}~\bibnamefont {Villing}}, \bibinfo {author} {\bibfnamefont
  {F.}~\bibnamefont {Moron}}, \bibinfo {author} {\bibfnamefont
  {M.}~\bibnamefont {Lours}}, \bibinfo {author} {\bibfnamefont
  {Y.}~\bibnamefont {Bidel}}, \bibinfo {author} {\bibfnamefont
  {A.}~\bibnamefont {Bresson}}, \bibinfo {author} {\bibfnamefont
  {A.}~\bibnamefont {Landragin}}, \ and\ \bibinfo {author} {\bibfnamefont
  {P.}~\bibnamefont {Bouyer}},\ }\href {\doibase 10.1038/ncomms1479} {\bibfield
   {journal} {\bibinfo  {journal} {Nature Communications}\ }\textbf {\bibinfo
  {volume} {2}},\ \bibinfo {pages} {474} (\bibinfo {year} {2011})}\BibitemShut
  {NoStop}%
\bibitem [{\citenamefont {{Cheiney}}\ \emph {et~al.}(2018)\citenamefont
  {{Cheiney}}, \citenamefont {{Fouch{\'e}}}, \citenamefont {{Templier}},
  \citenamefont {{Napolitano}}, \citenamefont {{Battelier}}, \citenamefont
  {{Bouyer}},\ and\ \citenamefont {{Barrett}}}]{Cheiney2018}%
  \BibitemOpen
  \bibfield  {author} {\bibinfo {author} {\bibfnamefont {P.}~\bibnamefont
  {{Cheiney}}}, \bibinfo {author} {\bibfnamefont {L.}~\bibnamefont
  {{Fouch{\'e}}}}, \bibinfo {author} {\bibfnamefont {S.}~\bibnamefont
  {{Templier}}}, \bibinfo {author} {\bibfnamefont {F.}~\bibnamefont
  {{Napolitano}}}, \bibinfo {author} {\bibfnamefont {B.}~\bibnamefont
  {{Battelier}}}, \bibinfo {author} {\bibfnamefont {P.}~\bibnamefont
  {{Bouyer}}}, \ and\ \bibinfo {author} {\bibfnamefont {B.}~\bibnamefont
  {{Barrett}}},\ }\href@noop {} {\bibfield  {journal} {\bibinfo  {journal}
  {ArXiv e-prints}\ } (\bibinfo {year} {2018})},\ \Eprint
  {http://arxiv.org/abs/1805.06198} {arXiv:1805.06198 [physics.atom-ph]}
  \BibitemShut {NoStop}%
\bibitem [{\citenamefont {Bouchendira}\ \emph {et~al.}(2011)\citenamefont
  {Bouchendira}, \citenamefont {Clad\'e}, \citenamefont {Guellati-Khélifa},
  \citenamefont {Nez},\ and\ \citenamefont {Biraben}}]{Bouchendira2011}%
  \BibitemOpen
  \bibfield  {author} {\bibinfo {author} {\bibfnamefont {R.}~\bibnamefont
  {Bouchendira}}, \bibinfo {author} {\bibfnamefont {P.}~\bibnamefont
  {Clad\'e}}, \bibinfo {author} {\bibfnamefont {S.}~\bibnamefont
  {Guellati-Khélifa}}, \bibinfo {author} {\bibfnamefont {F.}~\bibnamefont
  {Nez}}, \ and\ \bibinfo {author} {\bibfnamefont {F.}~\bibnamefont
  {Biraben}},\ }\href {\doibase 10.1103/PhysRevLett.106.080801} {\bibfield
  {journal} {\bibinfo  {journal} {Physical Review Letters}\ }\textbf {\bibinfo
  {volume} {106}} (\bibinfo {year} {2011}),\
  10.1103/PhysRevLett.106.080801}\BibitemShut {NoStop}%
\bibitem [{\citenamefont {Rosi}\ \emph {et~al.}(2014)\citenamefont {Rosi},
  \citenamefont {Sorrentino}, \citenamefont {Cacciapuoti}, \citenamefont
  {Prevedelli},\ and\ \citenamefont {Tino}}]{Rosi2014}%
  \BibitemOpen
  \bibfield  {author} {\bibinfo {author} {\bibfnamefont {G.}~\bibnamefont
  {Rosi}}, \bibinfo {author} {\bibfnamefont {F.}~\bibnamefont {Sorrentino}},
  \bibinfo {author} {\bibfnamefont {L.}~\bibnamefont {Cacciapuoti}}, \bibinfo
  {author} {\bibfnamefont {M.}~\bibnamefont {Prevedelli}}, \ and\ \bibinfo
  {author} {\bibfnamefont {G.~M.}\ \bibnamefont {Tino}},\ }\href {\doibase
  10.1038/nature13433} {\bibfield  {journal} {\bibinfo  {journal} {Nature}\
  }\textbf {\bibinfo {volume} {510}},\ \bibinfo {pages} {518} (\bibinfo {year}
  {2014})}\BibitemShut {NoStop}%
\bibitem [{\citenamefont {Parker}\ \emph {et~al.}(2018)\citenamefont {Parker},
  \citenamefont {Yu}, \citenamefont {Zhong}, \citenamefont {Estey},\ and\
  \citenamefont {M{\"u}ller}}]{Parker2018}%
  \BibitemOpen
  \bibfield  {author} {\bibinfo {author} {\bibfnamefont {R.~H.}\ \bibnamefont
  {Parker}}, \bibinfo {author} {\bibfnamefont {C.}~\bibnamefont {Yu}}, \bibinfo
  {author} {\bibfnamefont {W.}~\bibnamefont {Zhong}}, \bibinfo {author}
  {\bibfnamefont {B.}~\bibnamefont {Estey}}, \ and\ \bibinfo {author}
  {\bibfnamefont {H.}~\bibnamefont {M{\"u}ller}},\ }\href {\doibase
  10.1126/science.aap7706} {\bibfield  {journal} {\bibinfo  {journal}
  {Science}\ }\textbf {\bibinfo {volume} {360}},\ \bibinfo {pages} {191}
  (\bibinfo {year} {2018})},\ \Eprint
  {http://arxiv.org/abs/http://science.sciencemag.org/content/360/6385/191.full.pdf}
  {http://science.sciencemag.org/content/360/6385/191.full.pdf} \BibitemShut
  {NoStop}%
\bibitem [{\citenamefont {Peters}, \citenamefont {Chung},\ and\ \citenamefont
  {Chu}(2001)}]{Peters2001}%
  \BibitemOpen
  \bibfield  {author} {\bibinfo {author} {\bibfnamefont {A.}~\bibnamefont
  {Peters}}, \bibinfo {author} {\bibfnamefont {K.~Y.}\ \bibnamefont {Chung}}, \
  and\ \bibinfo {author} {\bibfnamefont {S.}~\bibnamefont {Chu}},\ }\href
  {http://stacks.iop.org/0026-1394/38/i=1/a=4} {\bibfield  {journal} {\bibinfo
  {journal} {Metrologia}\ }\textbf {\bibinfo {volume} {38}},\ \bibinfo {pages}
  {25} (\bibinfo {year} {2001})}\BibitemShut {NoStop}%
\bibitem [{\citenamefont {Hu}\ \emph {et~al.}(2013)\citenamefont {Hu},
  \citenamefont {Sun}, \citenamefont {Duan}, \citenamefont {Zhou},
  \citenamefont {Chen}, \citenamefont {Zhan}, \citenamefont {Zhang},\ and\
  \citenamefont {Luo}}]{Hu2013}%
  \BibitemOpen
  \bibfield  {author} {\bibinfo {author} {\bibfnamefont {Z.-K.}\ \bibnamefont
  {Hu}}, \bibinfo {author} {\bibfnamefont {B.-L.}\ \bibnamefont {Sun}},
  \bibinfo {author} {\bibfnamefont {X.-C.}\ \bibnamefont {Duan}}, \bibinfo
  {author} {\bibfnamefont {M.-K.}\ \bibnamefont {Zhou}}, \bibinfo {author}
  {\bibfnamefont {L.-L.}\ \bibnamefont {Chen}}, \bibinfo {author}
  {\bibfnamefont {S.}~\bibnamefont {Zhan}}, \bibinfo {author} {\bibfnamefont
  {Q.-Z.}\ \bibnamefont {Zhang}}, \ and\ \bibinfo {author} {\bibfnamefont
  {J.}~\bibnamefont {Luo}},\ }\href {\doibase 10.1103/PhysRevA.88.043610}
  {\bibfield  {journal} {\bibinfo  {journal} {Phys. Rev. A}\ }\textbf {\bibinfo
  {volume} {88}},\ \bibinfo {pages} {043610} (\bibinfo {year}
  {2013})}\BibitemShut {NoStop}%
\bibitem [{\citenamefont {Gillot}\ \emph {et~al.}(2014)\citenamefont {Gillot},
  \citenamefont {Francis}, \citenamefont {Landragin}, \citenamefont {Santos},\
  and\ \citenamefont {Merlet}}]{Gillot2014}%
  \BibitemOpen
  \bibfield  {author} {\bibinfo {author} {\bibfnamefont {P.}~\bibnamefont
  {Gillot}}, \bibinfo {author} {\bibfnamefont {O.}~\bibnamefont {Francis}},
  \bibinfo {author} {\bibfnamefont {A.}~\bibnamefont {Landragin}}, \bibinfo
  {author} {\bibfnamefont {F.~P.~D.}\ \bibnamefont {Santos}}, \ and\ \bibinfo
  {author} {\bibfnamefont {S.}~\bibnamefont {Merlet}},\ }\href
  {http://stacks.iop.org/0026-1394/51/i=5/a=L15} {\bibfield  {journal}
  {\bibinfo  {journal} {Metrologia}\ }\textbf {\bibinfo {volume} {51}},\
  \bibinfo {pages} {L15} (\bibinfo {year} {2014})}\BibitemShut {NoStop}%
\bibitem [{\citenamefont {Freier}\ \emph {et~al.}(2016)\citenamefont {Freier},
  \citenamefont {Hauth}, \citenamefont {Schkolnik}, \citenamefont {Leykauf},
  \citenamefont {Schilling}, \citenamefont {Wziontek}, \citenamefont
  {Scherneck}, \citenamefont {M\"{u}ller},\ and\ \citenamefont
  {Peters}}]{Freier2016}%
  \BibitemOpen
  \bibfield  {author} {\bibinfo {author} {\bibfnamefont {C.}~\bibnamefont
  {Freier}}, \bibinfo {author} {\bibfnamefont {M.}~\bibnamefont {Hauth}},
  \bibinfo {author} {\bibfnamefont {V.}~\bibnamefont {Schkolnik}}, \bibinfo
  {author} {\bibfnamefont {B.}~\bibnamefont {Leykauf}}, \bibinfo {author}
  {\bibfnamefont {M.}~\bibnamefont {Schilling}}, \bibinfo {author}
  {\bibfnamefont {H.}~\bibnamefont {Wziontek}}, \bibinfo {author}
  {\bibfnamefont {H.-G.}\ \bibnamefont {Scherneck}}, \bibinfo {author}
  {\bibfnamefont {J.}~\bibnamefont {M\"{u}ller}}, \ and\ \bibinfo {author}
  {\bibfnamefont {A.}~\bibnamefont {Peters}},\ }\href {\doibase
  10.1088/1742-6596/723/1/012050} {\bibfield  {journal} {\bibinfo  {journal}
  {J. Phys. Conf. Ser.}\ }\textbf {\bibinfo {volume} {723}},\ \bibinfo {pages}
  {012050} (\bibinfo {year} {2016})}\BibitemShut {NoStop}%
\bibitem [{\citenamefont {Wang}\ \emph {et~al.}(2018)\citenamefont {Wang},
  \citenamefont {Zhao}, \citenamefont {Zhuang}, \citenamefont {Li},
  \citenamefont {Wu}, \citenamefont {Feng},\ and\ \citenamefont
  {Li}}]{Wang2018}%
  \BibitemOpen
  \bibfield  {author} {\bibinfo {author} {\bibfnamefont {S.-K.}\ \bibnamefont
  {Wang}}, \bibinfo {author} {\bibfnamefont {Y.}~\bibnamefont {Zhao}}, \bibinfo
  {author} {\bibfnamefont {W.}~\bibnamefont {Zhuang}}, \bibinfo {author}
  {\bibfnamefont {T.-C.}\ \bibnamefont {Li}}, \bibinfo {author} {\bibfnamefont
  {S.-Q.}\ \bibnamefont {Wu}}, \bibinfo {author} {\bibfnamefont {J.-Y.}\
  \bibnamefont {Feng}}, \ and\ \bibinfo {author} {\bibfnamefont {C.-J.}\
  \bibnamefont {Li}},\ }\href {http://stacks.iop.org/0026-1394/55/i=3/a=360}
  {\bibfield  {journal} {\bibinfo  {journal} {Metrologia}\ }\textbf {\bibinfo
  {volume} {55}},\ \bibinfo {pages} {360} (\bibinfo {year} {2018})}\BibitemShut
  {NoStop}%
\bibitem [{\citenamefont {Bidel}\ \emph {et~al.}(2018)\citenamefont {Bidel},
  \citenamefont {Zahzam}, \citenamefont {Blanchard}, \citenamefont {Bonnin},
  \citenamefont {Cadoret}, \citenamefont {Bresson}, \citenamefont {Rouxel},\
  and\ \citenamefont {Lequentrec-Lalancette}}]{Bidel2018}%
  \BibitemOpen
  \bibfield  {author} {\bibinfo {author} {\bibfnamefont {Y.}~\bibnamefont
  {Bidel}}, \bibinfo {author} {\bibfnamefont {N.}~\bibnamefont {Zahzam}},
  \bibinfo {author} {\bibfnamefont {C.}~\bibnamefont {Blanchard}}, \bibinfo
  {author} {\bibfnamefont {A.}~\bibnamefont {Bonnin}}, \bibinfo {author}
  {\bibfnamefont {M.}~\bibnamefont {Cadoret}}, \bibinfo {author} {\bibfnamefont
  {A.}~\bibnamefont {Bresson}}, \bibinfo {author} {\bibfnamefont
  {D.}~\bibnamefont {Rouxel}}, \ and\ \bibinfo {author} {\bibfnamefont {M.~F.}\
  \bibnamefont {Lequentrec-Lalancette}},\ }\href
  {https://doi.org/10.1038/s41467-018-03040-2} {\bibfield  {journal} {\bibinfo
  {journal} {Nature Communications}\ }\textbf {\bibinfo {volume} {9}},\
  \bibinfo {pages} {627} (\bibinfo {year} {2018})}\BibitemShut {NoStop}%
\bibitem [{\citenamefont {{Karcher}}\ \emph {et~al.}(2018)\citenamefont
  {{Karcher}}, \citenamefont {{Imanaliev}}, \citenamefont {{Merlet}},\ and\
  \citenamefont {{Pereira dos Santos}}}]{Karcher2018}%
  \BibitemOpen
  \bibfield  {author} {\bibinfo {author} {\bibfnamefont {R.}~\bibnamefont
  {{Karcher}}}, \bibinfo {author} {\bibfnamefont {A.}~\bibnamefont
  {{Imanaliev}}}, \bibinfo {author} {\bibfnamefont {S.}~\bibnamefont
  {{Merlet}}}, \ and\ \bibinfo {author} {\bibfnamefont {F.}~\bibnamefont
  {{Pereira dos Santos}}},\ }\href {https://arxiv.org/abs/1804.04909}
  {\bibfield  {journal} {\bibinfo  {journal} {ArXiv e-prints}\ } (\bibinfo
  {year} {2018})},\ \Eprint {http://arxiv.org/abs/1804.04909} {arXiv:1804.04909
  [physics.atom-ph]} \BibitemShut {NoStop}%
\bibitem [{\citenamefont {McGuirk}\ \emph {et~al.}(2002)\citenamefont
  {McGuirk}, \citenamefont {Foster}, \citenamefont {Fixler}, \citenamefont
  {Snadden},\ and\ \citenamefont {Kasevich}}]{McGuirk2002}%
  \BibitemOpen
  \bibfield  {author} {\bibinfo {author} {\bibfnamefont {J.~M.}\ \bibnamefont
  {McGuirk}}, \bibinfo {author} {\bibfnamefont {G.~T.}\ \bibnamefont {Foster}},
  \bibinfo {author} {\bibfnamefont {J.~B.}\ \bibnamefont {Fixler}}, \bibinfo
  {author} {\bibfnamefont {M.~J.}\ \bibnamefont {Snadden}}, \ and\ \bibinfo
  {author} {\bibfnamefont {M.~A.}\ \bibnamefont {Kasevich}},\ }\href {\doibase
  10.1103/PhysRevA.65.033608} {\bibfield  {journal} {\bibinfo  {journal}
  {Physical Review A}\ }\textbf {\bibinfo {volume} {65}} (\bibinfo {year}
  {2002}),\ 10.1103/PhysRevA.65.033608}\BibitemShut {NoStop}%
\bibitem [{\citenamefont {Sorrentino}\ \emph {et~al.}(2012)\citenamefont
  {Sorrentino}, \citenamefont {Bertoldi}, \citenamefont {Bodart}, \citenamefont
  {Cacciapuoti}, \citenamefont {de~Angelis}, \citenamefont {Lien},
  \citenamefont {Prevedelli}, \citenamefont {Rosi},\ and\ \citenamefont
  {Tino}}]{Sorrentino2012}%
  \BibitemOpen
  \bibfield  {author} {\bibinfo {author} {\bibfnamefont {F.}~\bibnamefont
  {Sorrentino}}, \bibinfo {author} {\bibfnamefont {A.}~\bibnamefont
  {Bertoldi}}, \bibinfo {author} {\bibfnamefont {Q.}~\bibnamefont {Bodart}},
  \bibinfo {author} {\bibfnamefont {L.}~\bibnamefont {Cacciapuoti}}, \bibinfo
  {author} {\bibfnamefont {M.}~\bibnamefont {de~Angelis}}, \bibinfo {author}
  {\bibfnamefont {Y.-H.}\ \bibnamefont {Lien}}, \bibinfo {author}
  {\bibfnamefont {M.}~\bibnamefont {Prevedelli}}, \bibinfo {author}
  {\bibfnamefont {G.}~\bibnamefont {Rosi}}, \ and\ \bibinfo {author}
  {\bibfnamefont {G.~M.}\ \bibnamefont {Tino}},\ }\href {\doibase
  10.1063/1.4751112} {\bibfield  {journal} {\bibinfo  {journal} {Applied
  Physics Letters}\ }\textbf {\bibinfo {volume} {101}},\ \bibinfo {pages}
  {114106} (\bibinfo {year} {2012})}\BibitemShut {NoStop}%
\bibitem [{\citenamefont {Dimopoulos}\ \emph {et~al.}(2007)\citenamefont
  {Dimopoulos}, \citenamefont {Graham}, \citenamefont {Hogan},\ and\
  \citenamefont {Kasevich}}]{Dimopoulos2007}%
  \BibitemOpen
  \bibfield  {author} {\bibinfo {author} {\bibfnamefont {S.}~\bibnamefont
  {Dimopoulos}}, \bibinfo {author} {\bibfnamefont {P.~W.}\ \bibnamefont
  {Graham}}, \bibinfo {author} {\bibfnamefont {J.~M.}\ \bibnamefont {Hogan}}, \
  and\ \bibinfo {author} {\bibfnamefont {M.~A.}\ \bibnamefont {Kasevich}},\
  }\href {\doibase 10.1103/PhysRevLett.98.111102} {\bibfield  {journal}
  {\bibinfo  {journal} {Phys. Rev. Lett.}\ }\textbf {\bibinfo {volume} {98}},\
  \bibinfo {pages} {111102} (\bibinfo {year} {2007})}\BibitemShut {NoStop}%
\bibitem [{\citenamefont {Lepoutre}\ \emph {et~al.}(2012)\citenamefont
  {Lepoutre}, \citenamefont {Gauguet}, \citenamefont {Tr\'enec}, \citenamefont
  {B\"uchner},\ and\ \citenamefont {Vigu\'e}}]{Lepoutre2012}%
  \BibitemOpen
  \bibfield  {author} {\bibinfo {author} {\bibfnamefont {S.}~\bibnamefont
  {Lepoutre}}, \bibinfo {author} {\bibfnamefont {A.}~\bibnamefont {Gauguet}},
  \bibinfo {author} {\bibfnamefont {G.}~\bibnamefont {Tr\'enec}}, \bibinfo
  {author} {\bibfnamefont {M.}~\bibnamefont {B\"uchner}}, \ and\ \bibinfo
  {author} {\bibfnamefont {J.}~\bibnamefont {Vigu\'e}},\ }\href {\doibase
  10.1103/PhysRevLett.109.120404} {\bibfield  {journal} {\bibinfo  {journal}
  {Phys. Rev. Lett.}\ }\textbf {\bibinfo {volume} {109}},\ \bibinfo {pages}
  {120404} (\bibinfo {year} {2012})}\BibitemShut {NoStop}%
\bibitem [{\citenamefont {Aguilera}\ \emph {et~al.}(2014)\citenamefont
  {Aguilera}, \citenamefont {Ahlers}, \citenamefont {Battelier}, \citenamefont
  {Bawamia}, \citenamefont {Bertoldi}, \citenamefont {Bondarescu},
  \citenamefont {Bongs}, \citenamefont {Bouyer}, \citenamefont {Braxmaier},
  \citenamefont {Cacciapuoti},\ and\ \citenamefont {{et al}}}]{Aguilera2014}%
  \BibitemOpen
  \bibfield  {author} {\bibinfo {author} {\bibfnamefont {D.~N.}\ \bibnamefont
  {Aguilera}}, \bibinfo {author} {\bibfnamefont {H.}~\bibnamefont {Ahlers}},
  \bibinfo {author} {\bibfnamefont {B.}~\bibnamefont {Battelier}}, \bibinfo
  {author} {\bibfnamefont {A.}~\bibnamefont {Bawamia}}, \bibinfo {author}
  {\bibfnamefont {A.}~\bibnamefont {Bertoldi}}, \bibinfo {author}
  {\bibfnamefont {R.}~\bibnamefont {Bondarescu}}, \bibinfo {author}
  {\bibfnamefont {K.}~\bibnamefont {Bongs}}, \bibinfo {author} {\bibfnamefont
  {P.}~\bibnamefont {Bouyer}}, \bibinfo {author} {\bibfnamefont
  {C.}~\bibnamefont {Braxmaier}}, \bibinfo {author} {\bibfnamefont
  {L.}~\bibnamefont {Cacciapuoti}}, \ and\ \bibinfo {author} {\bibnamefont {{et
  al}}},\ }\href {http://stacks.iop.org/0264-9381/31/i=11/a=115010} {\bibfield
  {journal} {\bibinfo  {journal} {Classical and Quantum Gravity}\ }\textbf
  {\bibinfo {volume} {31}},\ \bibinfo {pages} {115010} (\bibinfo {year}
  {2014})}\BibitemShut {NoStop}%
\bibitem [{\citenamefont {Zhou}\ \emph {et~al.}(2015)\citenamefont {Zhou},
  \citenamefont {Long}, \citenamefont {Tang}, \citenamefont {Chen},
  \citenamefont {Gao}, \citenamefont {Peng}, \citenamefont {Duan},
  \citenamefont {Zhong}, \citenamefont {Xiong}, \citenamefont {Wang},
  \citenamefont {Zhang},\ and\ \citenamefont {Zhan}}]{Zhou2015}%
  \BibitemOpen
  \bibfield  {author} {\bibinfo {author} {\bibfnamefont {L.}~\bibnamefont
  {Zhou}}, \bibinfo {author} {\bibfnamefont {S.}~\bibnamefont {Long}}, \bibinfo
  {author} {\bibfnamefont {B.}~\bibnamefont {Tang}}, \bibinfo {author}
  {\bibfnamefont {X.}~\bibnamefont {Chen}}, \bibinfo {author} {\bibfnamefont
  {F.}~\bibnamefont {Gao}}, \bibinfo {author} {\bibfnamefont {W.}~\bibnamefont
  {Peng}}, \bibinfo {author} {\bibfnamefont {W.}~\bibnamefont {Duan}}, \bibinfo
  {author} {\bibfnamefont {J.}~\bibnamefont {Zhong}}, \bibinfo {author}
  {\bibfnamefont {Z.}~\bibnamefont {Xiong}}, \bibinfo {author} {\bibfnamefont
  {J.}~\bibnamefont {Wang}}, \bibinfo {author} {\bibfnamefont {Y.}~\bibnamefont
  {Zhang}}, \ and\ \bibinfo {author} {\bibfnamefont {M.}~\bibnamefont {Zhan}},\
  }\href {\doibase 10.1103/PhysRevLett.115.013004} {\bibfield  {journal}
  {\bibinfo  {journal} {Physical Review Letters}\ }\textbf {\bibinfo {volume}
  {115}} (\bibinfo {year} {2015}),\ 10.1103/PhysRevLett.115.013004}\BibitemShut
  {NoStop}%
\bibitem [{\citenamefont {Jaffe}\ \emph {et~al.}(2017)\citenamefont {Jaffe},
  \citenamefont {Haslinger}, \citenamefont {Xu}, \citenamefont {Hamilton},
  \citenamefont {Upadhye}, \citenamefont {Elder}, \citenamefont {Khoury},\ and\
  \citenamefont {Müller}}]{Jaffe2017}%
  \BibitemOpen
  \bibfield  {author} {\bibinfo {author} {\bibfnamefont {M.}~\bibnamefont
  {Jaffe}}, \bibinfo {author} {\bibfnamefont {P.}~\bibnamefont {Haslinger}},
  \bibinfo {author} {\bibfnamefont {V.}~\bibnamefont {Xu}}, \bibinfo {author}
  {\bibfnamefont {P.}~\bibnamefont {Hamilton}}, \bibinfo {author}
  {\bibfnamefont {A.}~\bibnamefont {Upadhye}}, \bibinfo {author} {\bibfnamefont
  {B.}~\bibnamefont {Elder}}, \bibinfo {author} {\bibfnamefont
  {J.}~\bibnamefont {Khoury}}, \ and\ \bibinfo {author} {\bibfnamefont
  {H.}~\bibnamefont {Müller}},\ }\href {\doibase 10.1038/nphys4189} {\bibfield
   {journal} {\bibinfo  {journal} {Nature Physics}\ }\textbf {\bibinfo {volume}
  {13}},\ \bibinfo {pages} {938} (\bibinfo {year} {2017})}\BibitemShut
  {NoStop}%
\bibitem [{\citenamefont {Asenbaum}\ \emph {et~al.}(2017)\citenamefont
  {Asenbaum}, \citenamefont {Overstreet}, \citenamefont {Kovachy},
  \citenamefont {Brown}, \citenamefont {Hogan},\ and\ \citenamefont
  {Kasevich}}]{Asenbaum2017}%
  \BibitemOpen
  \bibfield  {author} {\bibinfo {author} {\bibfnamefont {P.}~\bibnamefont
  {Asenbaum}}, \bibinfo {author} {\bibfnamefont {C.}~\bibnamefont
  {Overstreet}}, \bibinfo {author} {\bibfnamefont {T.}~\bibnamefont {Kovachy}},
  \bibinfo {author} {\bibfnamefont {D.~D.}\ \bibnamefont {Brown}}, \bibinfo
  {author} {\bibfnamefont {J.~M.}\ \bibnamefont {Hogan}}, \ and\ \bibinfo
  {author} {\bibfnamefont {M.~A.}\ \bibnamefont {Kasevich}},\ }\href {\doibase
  10.1103/PhysRevLett.118.183602} {\bibfield  {journal} {\bibinfo  {journal}
  {Physical Review Letters}\ }\textbf {\bibinfo {volume} {118}} (\bibinfo
  {year} {2017}),\ 10.1103/PhysRevLett.118.183602}\BibitemShut {NoStop}%
\bibitem [{\citenamefont {Chaibi}\ \emph {et~al.}(2016)\citenamefont {Chaibi},
  \citenamefont {Geiger}, \citenamefont {Canuel}, \citenamefont {Bertoldi},
  \citenamefont {Landragin},\ and\ \citenamefont {Bouyer}}]{Chaibi2016}%
  \BibitemOpen
  \bibfield  {author} {\bibinfo {author} {\bibfnamefont {W.}~\bibnamefont
  {Chaibi}}, \bibinfo {author} {\bibfnamefont {R.}~\bibnamefont {Geiger}},
  \bibinfo {author} {\bibfnamefont {B.}~\bibnamefont {Canuel}}, \bibinfo
  {author} {\bibfnamefont {A.}~\bibnamefont {Bertoldi}}, \bibinfo {author}
  {\bibfnamefont {A.}~\bibnamefont {Landragin}}, \ and\ \bibinfo {author}
  {\bibfnamefont {P.}~\bibnamefont {Bouyer}},\ }\href {\doibase
  10.1103/PhysRevD.93.021101} {\bibfield  {journal} {\bibinfo  {journal}
  {Physical Review D}\ }\textbf {\bibinfo {volume} {93}} (\bibinfo {year}
  {2016}),\ 10.1103/PhysRevD.93.021101}\BibitemShut {NoStop}%
\bibitem [{\citenamefont {Hogan}\ and\ \citenamefont
  {Kasevich}(2016)}]{Hogan2016}%
  \BibitemOpen
  \bibfield  {author} {\bibinfo {author} {\bibfnamefont {J.~M.}\ \bibnamefont
  {Hogan}}\ and\ \bibinfo {author} {\bibfnamefont {M.~A.}\ \bibnamefont
  {Kasevich}},\ }\href {\doibase 10.1103/PhysRevA.94.033632} {\bibfield
  {journal} {\bibinfo  {journal} {Phys. Rev. A}\ }\textbf {\bibinfo {volume}
  {94}},\ \bibinfo {pages} {033632} (\bibinfo {year} {2016})}\BibitemShut
  {NoStop}%
\bibitem [{\citenamefont {Dutta}\ \emph {et~al.}(2016)\citenamefont {Dutta},
  \citenamefont {Savoie}, \citenamefont {Fang}, \citenamefont {Venon},
  \citenamefont {Garrido~Alzar}, \citenamefont {Geiger},\ and\ \citenamefont
  {Landragin}}]{Dutta2016}%
  \BibitemOpen
  \bibfield  {author} {\bibinfo {author} {\bibfnamefont {I.}~\bibnamefont
  {Dutta}}, \bibinfo {author} {\bibfnamefont {D.}~\bibnamefont {Savoie}},
  \bibinfo {author} {\bibfnamefont {B.}~\bibnamefont {Fang}}, \bibinfo {author}
  {\bibfnamefont {B.}~\bibnamefont {Venon}}, \bibinfo {author} {\bibfnamefont
  {C.}~\bibnamefont {Garrido~Alzar}}, \bibinfo {author} {\bibfnamefont
  {R.}~\bibnamefont {Geiger}}, \ and\ \bibinfo {author} {\bibfnamefont
  {A.}~\bibnamefont {Landragin}},\ }\href {\doibase
  10.1103/PhysRevLett.116.183003} {\bibfield  {journal} {\bibinfo  {journal}
  {Physical Review Letters}\ }\textbf {\bibinfo {volume} {116}} (\bibinfo
  {year} {2016}),\ 10.1103/PhysRevLett.116.183003}\BibitemShut {NoStop}%
\bibitem [{\citenamefont {Gauguet}\ \emph {et~al.}(2009)\citenamefont
  {Gauguet}, \citenamefont {Canuel}, \citenamefont {L\'ev\`eque}, \citenamefont
  {Chaibi},\ and\ \citenamefont {Landragin}}]{Gauguet2009}%
  \BibitemOpen
  \bibfield  {author} {\bibinfo {author} {\bibfnamefont {A.}~\bibnamefont
  {Gauguet}}, \bibinfo {author} {\bibfnamefont {B.}~\bibnamefont {Canuel}},
  \bibinfo {author} {\bibfnamefont {T.}~\bibnamefont {L\'ev\`eque}}, \bibinfo
  {author} {\bibfnamefont {W.}~\bibnamefont {Chaibi}}, \ and\ \bibinfo {author}
  {\bibfnamefont {A.}~\bibnamefont {Landragin}},\ }\href {\doibase
  10.1103/PhysRevA.80.063604} {\bibfield  {journal} {\bibinfo  {journal} {Phys.
  Rev. A}\ }\textbf {\bibinfo {volume} {80}},\ \bibinfo {pages} {063604}
  (\bibinfo {year} {2009})}\BibitemShut {NoStop}%
\bibitem [{\citenamefont {Mazzoni}\ \emph {et~al.}(2015)\citenamefont
  {Mazzoni}, \citenamefont {Zhang}, \citenamefont {Del~Aguila}, \citenamefont
  {Salvi}, \citenamefont {Poli},\ and\ \citenamefont {Tino}}]{Mazzoni2015}%
  \BibitemOpen
  \bibfield  {author} {\bibinfo {author} {\bibfnamefont {T.}~\bibnamefont
  {Mazzoni}}, \bibinfo {author} {\bibfnamefont {X.}~\bibnamefont {Zhang}},
  \bibinfo {author} {\bibfnamefont {R.}~\bibnamefont {Del~Aguila}}, \bibinfo
  {author} {\bibfnamefont {L.}~\bibnamefont {Salvi}}, \bibinfo {author}
  {\bibfnamefont {N.}~\bibnamefont {Poli}}, \ and\ \bibinfo {author}
  {\bibfnamefont {G.~M.}\ \bibnamefont {Tino}},\ }\href {\doibase
  10.1103/PhysRevA.92.053619} {\bibfield  {journal} {\bibinfo  {journal} {Phys.
  Rev. A}\ }\textbf {\bibinfo {volume} {92}},\ \bibinfo {pages} {053619}
  (\bibinfo {year} {2015})}\BibitemShut {NoStop}%
\bibitem [{\citenamefont {B\"uchner}\ \emph {et~al.}(2003)\citenamefont
  {B\"uchner}, \citenamefont {Delhuille}, \citenamefont {Miffre}, \citenamefont
  {Robilliard}, \citenamefont {Vigu\'e},\ and\ \citenamefont
  {Champenois}}]{Buechner2003}%
  \BibitemOpen
  \bibfield  {author} {\bibinfo {author} {\bibfnamefont {M.}~\bibnamefont
  {B\"uchner}}, \bibinfo {author} {\bibfnamefont {R.}~\bibnamefont
  {Delhuille}}, \bibinfo {author} {\bibfnamefont {A.}~\bibnamefont {Miffre}},
  \bibinfo {author} {\bibfnamefont {C.}~\bibnamefont {Robilliard}}, \bibinfo
  {author} {\bibfnamefont {J.}~\bibnamefont {Vigu\'e}}, \ and\ \bibinfo
  {author} {\bibfnamefont {C.}~\bibnamefont {Champenois}},\ }\href {\doibase
  10.1103/PhysRevA.68.013607} {\bibfield  {journal} {\bibinfo  {journal} {Phys.
  Rev. A}\ }\textbf {\bibinfo {volume} {68}},\ \bibinfo {pages} {013607}
  (\bibinfo {year} {2003})}\BibitemShut {NoStop}%
\bibitem [{\citenamefont {Gori}(1994)}]{Gori1994}%
  \BibitemOpen
  \bibfield  {author} {\bibinfo {author} {\bibfnamefont {F.}~\bibnamefont
  {Gori}},\ }\href
  {http://www.sciencedirect.com/science/article/pii/0030401894903425}
  {\bibfield  {journal} {\bibinfo  {journal} {Optics Communications}\ }\textbf
  {\bibinfo {volume} {107}},\ \bibinfo {pages} {335} (\bibinfo {year}
  {1994})}\BibitemShut {NoStop}%
\bibitem [{\citenamefont {Pal}\ \emph {et~al.}(2018)\citenamefont {Pal},
  \citenamefont {Tradonsky}, \citenamefont {Chriki}, \citenamefont {Kaplan},
  \citenamefont {Brodsky}, \citenamefont {Attia}, \citenamefont {Davidson},\
  and\ \citenamefont {Friesem}}]{Pal2018}%
  \BibitemOpen
  \bibfield  {author} {\bibinfo {author} {\bibfnamefont {V.}~\bibnamefont
  {Pal}}, \bibinfo {author} {\bibfnamefont {C.}~\bibnamefont {Tradonsky}},
  \bibinfo {author} {\bibfnamefont {R.}~\bibnamefont {Chriki}}, \bibinfo
  {author} {\bibfnamefont {N.}~\bibnamefont {Kaplan}}, \bibinfo {author}
  {\bibfnamefont {A.}~\bibnamefont {Brodsky}}, \bibinfo {author} {\bibfnamefont
  {M.}~\bibnamefont {Attia}}, \bibinfo {author} {\bibfnamefont
  {N.}~\bibnamefont {Davidson}}, \ and\ \bibinfo {author} {\bibfnamefont
  {A.~A.}\ \bibnamefont {Friesem}},\ }\href {\doibase 10.1364/AO.57.004583}
  {\bibfield  {journal} {\bibinfo  {journal} {Appl. Opt.}\ }\textbf {\bibinfo
  {volume} {57}},\ \bibinfo {pages} {4583} (\bibinfo {year}
  {2018})}\BibitemShut {NoStop}%
\bibitem [{\citenamefont {Ma}\ \emph {et~al.}(2010)\citenamefont {Ma},
  \citenamefont {Liu}, \citenamefont {Zhou}, \citenamefont {Wang},
  \citenamefont {Ma},\ and\ \citenamefont {Xu}}]{Ma2010}%
  \BibitemOpen
  \bibfield  {author} {\bibinfo {author} {\bibfnamefont {H.}~\bibnamefont
  {Ma}}, \bibinfo {author} {\bibfnamefont {Z.}~\bibnamefont {Liu}}, \bibinfo
  {author} {\bibfnamefont {P.}~\bibnamefont {Zhou}}, \bibinfo {author}
  {\bibfnamefont {X.}~\bibnamefont {Wang}}, \bibinfo {author} {\bibfnamefont
  {Y.}~\bibnamefont {Ma}}, \ and\ \bibinfo {author} {\bibfnamefont
  {X.}~\bibnamefont {Xu}},\ }\href
  {http://stacks.iop.org/2040-8986/12/i=4/a=045704} {\bibfield  {journal}
  {\bibinfo  {journal} {Journal of Optics}\ }\textbf {\bibinfo {volume} {12}},\
  \bibinfo {pages} {045704} (\bibinfo {year} {2010})}\BibitemShut {NoStop}%
\bibitem [{\citenamefont {Hoffnagle}\ and\ \citenamefont
  {Jefferson}(2000)}]{Hoffnagle2000}%
  \BibitemOpen
  \bibfield  {author} {\bibinfo {author} {\bibfnamefont {J.~A.}\ \bibnamefont
  {Hoffnagle}}\ and\ \bibinfo {author} {\bibfnamefont {C.~M.}\ \bibnamefont
  {Jefferson}},\ }\href {\doibase 10.1364/AO.39.005488} {\bibfield  {journal}
  {\bibinfo  {journal} {Appl. Opt.}\ }\textbf {\bibinfo {volume} {39}},\
  \bibinfo {pages} {5488} (\bibinfo {year} {2000})}\BibitemShut {NoStop}%
\bibitem [{asp()}]{asphericonWebsite}%
  \BibitemOpen
  \href
  {https://www.asphericon.com/en/asphere/shape-it-til-you-make-it-top-hat-beam-shaping-with-aspheres/}
  {}\bibinfo {note} {Asphericon website, cited \today}\BibitemShut {NoStop}%
\bibitem [{\citenamefont {Louchet-Chauvet}\ \emph {et~al.}(2011)\citenamefont
  {Louchet-Chauvet}, \citenamefont {Farah}, \citenamefont {Bodart},
  \citenamefont {Clairon}, \citenamefont {Landragin}, \citenamefont {Merlet},\
  and\ \citenamefont {Santos}}]{Louchet-Chauvet2011}%
  \BibitemOpen
  \bibfield  {author} {\bibinfo {author} {\bibfnamefont {A.}~\bibnamefont
  {Louchet-Chauvet}}, \bibinfo {author} {\bibfnamefont {T.}~\bibnamefont
  {Farah}}, \bibinfo {author} {\bibfnamefont {Q.}~\bibnamefont {Bodart}},
  \bibinfo {author} {\bibfnamefont {A.}~\bibnamefont {Clairon}}, \bibinfo
  {author} {\bibfnamefont {A.}~\bibnamefont {Landragin}}, \bibinfo {author}
  {\bibfnamefont {S.}~\bibnamefont {Merlet}}, \ and\ \bibinfo {author}
  {\bibfnamefont {F.~P.~D.}\ \bibnamefont {Santos}},\ }\href
  {http://stacks.iop.org/1367-2630/13/i=6/a=065025} {\bibfield  {journal}
  {\bibinfo  {journal} {New Journal of Physics}\ }\textbf {\bibinfo {volume}
  {13}},\ \bibinfo {pages} {065025} (\bibinfo {year} {2011})}\BibitemShut
  {NoStop}%
\bibitem [{\citenamefont {Takeda}, \citenamefont {Ina},\ and\ \citenamefont
  {Kobayashi}()}]{Takeda1982}%
  \BibitemOpen
  \bibfield  {author} {\bibinfo {author} {\bibfnamefont {M.}~\bibnamefont
  {Takeda}}, \bibinfo {author} {\bibfnamefont {H.}~\bibnamefont {Ina}}, \ and\
  \bibinfo {author} {\bibfnamefont {S.}~\bibnamefont {Kobayashi}},\ }\href
  {\doibase 10.1364/JOSA.72.000156} {\ \textbf {\bibinfo {volume} {72}},\
  \bibinfo {pages} {156}}\BibitemShut {NoStop}%
\bibitem [{\citenamefont {Schkolnik}\ \emph {et~al.}(2015)\citenamefont
  {Schkolnik}, \citenamefont {Leykauf}, \citenamefont {Hauth}, \citenamefont
  {Freier},\ and\ \citenamefont {Peters}}]{Schkolnik2015}%
  \BibitemOpen
  \bibfield  {author} {\bibinfo {author} {\bibfnamefont {V.}~\bibnamefont
  {Schkolnik}}, \bibinfo {author} {\bibfnamefont {B.}~\bibnamefont {Leykauf}},
  \bibinfo {author} {\bibfnamefont {M.}~\bibnamefont {Hauth}}, \bibinfo
  {author} {\bibfnamefont {C.}~\bibnamefont {Freier}}, \ and\ \bibinfo {author}
  {\bibfnamefont {A.}~\bibnamefont {Peters}},\ }\href
  {https://doi.org/10.1007/s00340-015-6138-5} {\bibfield  {journal} {\bibinfo
  {journal} {Applied Physics B}\ }\textbf {\bibinfo {volume} {120}},\ \bibinfo
  {pages} {311} (\bibinfo {year} {2015})}\BibitemShut {NoStop}%
\bibitem [{\citenamefont {Meunier}\ \emph {et~al.}(2014)\citenamefont
  {Meunier}, \citenamefont {Dutta}, \citenamefont {Geiger}, \citenamefont
  {Guerlin}, \citenamefont {Garrido~Alzar},\ and\ \citenamefont
  {Landragin}}]{Meunier2014}%
  \BibitemOpen
  \bibfield  {author} {\bibinfo {author} {\bibfnamefont {M.}~\bibnamefont
  {Meunier}}, \bibinfo {author} {\bibfnamefont {I.}~\bibnamefont {Dutta}},
  \bibinfo {author} {\bibfnamefont {R.}~\bibnamefont {Geiger}}, \bibinfo
  {author} {\bibfnamefont {C.}~\bibnamefont {Guerlin}}, \bibinfo {author}
  {\bibfnamefont {C.~L.}\ \bibnamefont {Garrido~Alzar}}, \ and\ \bibinfo
  {author} {\bibfnamefont {A.}~\bibnamefont {Landragin}},\ }\href {\doibase
  10.1103/PhysRevA.90.063633} {\bibfield  {journal} {\bibinfo  {journal}
  {Physical Review A}\ }\textbf {\bibinfo {volume} {90}} (\bibinfo {year}
  {2014}),\ 10.1103/PhysRevA.90.063633}\BibitemShut {NoStop}%
\bibitem [{not()}]{noteSpatialFreq}%
  \BibitemOpen
  \href@noop {} {}\bibinfo {note} {We varied the spatial frequencies of the
  added intensity noise and found no substantial difference in the simulation
  results as long as the spatial period was smaller than about 1/10 of the beam
  size. In the simulation reported in Fig. 3, the spatial period of the noise
  is 1/100 of the beam size.}\BibitemShut {Stop}%
\bibitem [{\citenamefont {{Merlet}}\ \emph {et~al.}(2009)\citenamefont
  {{Merlet}}, \citenamefont {{LeGou{\"e}t}}, \citenamefont {{Bodart}},
  \citenamefont {{Clairon}}, \citenamefont {{Landragin}}, \citenamefont
  {{Pereira Dos Santos}},\ and\ \citenamefont {{Rouchon}}}]{Merlet2009}%
  \BibitemOpen
  \bibfield  {author} {\bibinfo {author} {\bibfnamefont {S.}~\bibnamefont
  {{Merlet}}}, \bibinfo {author} {\bibfnamefont {J.}~\bibnamefont
  {{LeGou{\"e}t}}}, \bibinfo {author} {\bibfnamefont {Q.}~\bibnamefont
  {{Bodart}}}, \bibinfo {author} {\bibfnamefont {A.}~\bibnamefont {{Clairon}}},
  \bibinfo {author} {\bibfnamefont {A.}~\bibnamefont {{Landragin}}}, \bibinfo
  {author} {\bibfnamefont {F.}~\bibnamefont {{Pereira Dos Santos}}}, \ and\
  \bibinfo {author} {\bibfnamefont {P.}~\bibnamefont {{Rouchon}}},\ }\href
  {\doibase 10.1088/0026-1394/46/1/011} {\bibfield  {journal} {\bibinfo
  {journal} {Metrologia}\ }\textbf {\bibinfo {volume} {46}},\ \bibinfo {pages}
  {87} (\bibinfo {year} {2009})},\ \Eprint {http://arxiv.org/abs/0806.0164}
  {arXiv:0806.0164 [physics.atom-ph]} \BibitemShut {NoStop}%
\bibitem [{\citenamefont {Gauguet}\ \emph {et~al.}(2008)\citenamefont
  {Gauguet}, \citenamefont {Mehlst\"aubler}, \citenamefont {L\'ev\`eque},
  \citenamefont {Le~Gou\"et}, \citenamefont {Chaibi}, \citenamefont {Canuel},
  \citenamefont {Clairon}, \citenamefont {Dos~Santos},\ and\ \citenamefont
  {Landragin}}]{Gauguet2008}%
  \BibitemOpen
  \bibfield  {author} {\bibinfo {author} {\bibfnamefont {A.}~\bibnamefont
  {Gauguet}}, \bibinfo {author} {\bibfnamefont {T.~E.}\ \bibnamefont
  {Mehlst\"aubler}}, \bibinfo {author} {\bibfnamefont {T.}~\bibnamefont
  {L\'ev\`eque}}, \bibinfo {author} {\bibfnamefont {J.}~\bibnamefont
  {Le~Gou\"et}}, \bibinfo {author} {\bibfnamefont {W.}~\bibnamefont {Chaibi}},
  \bibinfo {author} {\bibfnamefont {B.}~\bibnamefont {Canuel}}, \bibinfo
  {author} {\bibfnamefont {A.}~\bibnamefont {Clairon}}, \bibinfo {author}
  {\bibfnamefont {F.~P.}\ \bibnamefont {Dos~Santos}}, \ and\ \bibinfo {author}
  {\bibfnamefont {A.}~\bibnamefont {Landragin}},\ }\href {\doibase
  10.1103/PhysRevA.78.043615} {\bibfield  {journal} {\bibinfo  {journal} {Phys.
  Rev. A}\ }\textbf {\bibinfo {volume} {78}},\ \bibinfo {pages} {043615}
  (\bibinfo {year} {2008})}\BibitemShut {NoStop}%
\bibitem [{\citenamefont {Tackmann}\ \emph {et~al.}(2012)\citenamefont
  {Tackmann}, \citenamefont {Berg}, \citenamefont {Schubert}, \citenamefont
  {Abend}, \citenamefont {Gilowski}, \citenamefont {Ertmer},\ and\
  \citenamefont {Rasel}}]{Tackmann2012}%
  \BibitemOpen
  \bibfield  {author} {\bibinfo {author} {\bibfnamefont {G.}~\bibnamefont
  {Tackmann}}, \bibinfo {author} {\bibfnamefont {P.}~\bibnamefont {Berg}},
  \bibinfo {author} {\bibfnamefont {C.}~\bibnamefont {Schubert}}, \bibinfo
  {author} {\bibfnamefont {S.}~\bibnamefont {Abend}}, \bibinfo {author}
  {\bibfnamefont {M.}~\bibnamefont {Gilowski}}, \bibinfo {author}
  {\bibfnamefont {W.}~\bibnamefont {Ertmer}}, \ and\ \bibinfo {author}
  {\bibfnamefont {E.~M.}\ \bibnamefont {Rasel}},\ }\href
  {http://stacks.iop.org/1367-2630/14/i=1/a=015002} {\bibfield  {journal}
  {\bibinfo  {journal} {New Journal of Physics}\ }\textbf {\bibinfo {volume}
  {14}},\ \bibinfo {pages} {015002} (\bibinfo {year} {2012})}\BibitemShut
  {NoStop}%
\bibitem [{\citenamefont {Yao}\ \emph {et~al.}(2018)\citenamefont {Yao},
  \citenamefont {Lu}, \citenamefont {Li}, \citenamefont {Luo}, \citenamefont
  {Wang},\ and\ \citenamefont {Zhan}}]{Yao2018}%
  \BibitemOpen
  \bibfield  {author} {\bibinfo {author} {\bibfnamefont {Z.-W.}\ \bibnamefont
  {Yao}}, \bibinfo {author} {\bibfnamefont {S.-B.}\ \bibnamefont {Lu}},
  \bibinfo {author} {\bibfnamefont {R.-B.}\ \bibnamefont {Li}}, \bibinfo
  {author} {\bibfnamefont {J.}~\bibnamefont {Luo}}, \bibinfo {author}
  {\bibfnamefont {J.}~\bibnamefont {Wang}}, \ and\ \bibinfo {author}
  {\bibfnamefont {M.-S.}\ \bibnamefont {Zhan}},\ }\href {\doibase
  10.1103/PhysRevA.97.013620} {\bibfield  {journal} {\bibinfo  {journal}
  {Physical Review A}\ }\textbf {\bibinfo {volume} {97}} (\bibinfo {year}
  {2018}),\ 10.1103/PhysRevA.97.013620}\BibitemShut {NoStop}%
\bibitem [{\citenamefont {Chiow}\ \emph {et~al.}(2011)\citenamefont {Chiow},
  \citenamefont {Kovachy}, \citenamefont {Chien},\ and\ \citenamefont
  {Kasevich}}]{Chiow2011}%
  \BibitemOpen
  \bibfield  {author} {\bibinfo {author} {\bibfnamefont {S.-w.}\ \bibnamefont
  {Chiow}}, \bibinfo {author} {\bibfnamefont {T.}~\bibnamefont {Kovachy}},
  \bibinfo {author} {\bibfnamefont {H.-C.}\ \bibnamefont {Chien}}, \ and\
  \bibinfo {author} {\bibfnamefont {M.~A.}\ \bibnamefont {Kasevich}},\ }\href
  {\doibase 10.1103/PhysRevLett.107.130403} {\bibfield  {journal} {\bibinfo
  {journal} {Physical Review Letters}\ }\textbf {\bibinfo {volume} {107}}
  (\bibinfo {year} {2011}),\ 10.1103/PhysRevLett.107.130403}\BibitemShut
  {NoStop}%
\bibitem [{\citenamefont {Altin}\ \emph {et~al.}(2013)\citenamefont {Altin},
  \citenamefont {Johnsson}, \citenamefont {Negnevitsky}, \citenamefont
  {Dennis}, \citenamefont {Anderson}, \citenamefont {Debs}, \citenamefont
  {Szigeti}, \citenamefont {Hardman}, \citenamefont {Bennetts}, \citenamefont
  {McDonald}, \citenamefont {Turner}, \citenamefont {Close},\ and\
  \citenamefont {Robins}}]{Altin2013}%
  \BibitemOpen
  \bibfield  {author} {\bibinfo {author} {\bibfnamefont {P.~A.}\ \bibnamefont
  {Altin}}, \bibinfo {author} {\bibfnamefont {M.~T.}\ \bibnamefont {Johnsson}},
  \bibinfo {author} {\bibfnamefont {V.}~\bibnamefont {Negnevitsky}}, \bibinfo
  {author} {\bibfnamefont {G.~R.}\ \bibnamefont {Dennis}}, \bibinfo {author}
  {\bibfnamefont {R.~P.}\ \bibnamefont {Anderson}}, \bibinfo {author}
  {\bibfnamefont {J.~E.}\ \bibnamefont {Debs}}, \bibinfo {author}
  {\bibfnamefont {S.~S.}\ \bibnamefont {Szigeti}}, \bibinfo {author}
  {\bibfnamefont {K.~S.}\ \bibnamefont {Hardman}}, \bibinfo {author}
  {\bibfnamefont {S.}~\bibnamefont {Bennetts}}, \bibinfo {author}
  {\bibfnamefont {G.~D.}\ \bibnamefont {McDonald}}, \bibinfo {author}
  {\bibfnamefont {L.~D.}\ \bibnamefont {Turner}}, \bibinfo {author}
  {\bibfnamefont {J.~D.}\ \bibnamefont {Close}}, \ and\ \bibinfo {author}
  {\bibfnamefont {N.~P.}\ \bibnamefont {Robins}},\ }\href
  {http://stacks.iop.org/1367-2630/15/i=2/a=023009} {\bibfield  {journal}
  {\bibinfo  {journal} {New Journal of Physics}\ }\textbf {\bibinfo {volume}
  {15}},\ \bibinfo {pages} {023009} (\bibinfo {year} {2013})}\BibitemShut
  {NoStop}%
\bibitem [{\citenamefont {Estey}\ \emph {et~al.}(2015)\citenamefont {Estey},
  \citenamefont {Yu}, \citenamefont {M\"uller}, \citenamefont {Kuan},\ and\
  \citenamefont {Lan}}]{Estey2015}%
  \BibitemOpen
  \bibfield  {author} {\bibinfo {author} {\bibfnamefont {B.}~\bibnamefont
  {Estey}}, \bibinfo {author} {\bibfnamefont {C.}~\bibnamefont {Yu}}, \bibinfo
  {author} {\bibfnamefont {H.}~\bibnamefont {M\"uller}}, \bibinfo {author}
  {\bibfnamefont {P.-C.}\ \bibnamefont {Kuan}}, \ and\ \bibinfo {author}
  {\bibfnamefont {S.-Y.}\ \bibnamefont {Lan}},\ }\href {\doibase
  10.1103/PhysRevLett.115.083002} {\bibfield  {journal} {\bibinfo  {journal}
  {Phys. Rev. Lett.}\ }\textbf {\bibinfo {volume} {115}},\ \bibinfo {pages}
  {083002} (\bibinfo {year} {2015})}\BibitemShut {NoStop}%
\bibitem [{\citenamefont {Riou}\ \emph {et~al.}(2017)\citenamefont {Riou},
  \citenamefont {Mielec}, \citenamefont {Lefèvre}, \citenamefont {Prevedelli},
  \citenamefont {Landragin}, \citenamefont {Bouyer}, \citenamefont {Bertoldi},
  \citenamefont {Geiger},\ and\ \citenamefont {Canuel}}]{riou_marginally_2017}%
  \BibitemOpen
  \bibfield  {author} {\bibinfo {author} {\bibfnamefont {I.}~\bibnamefont
  {Riou}}, \bibinfo {author} {\bibfnamefont {N.}~\bibnamefont {Mielec}},
  \bibinfo {author} {\bibfnamefont {G.}~\bibnamefont {Lefèvre}}, \bibinfo
  {author} {\bibfnamefont {M.}~\bibnamefont {Prevedelli}}, \bibinfo {author}
  {\bibfnamefont {A.}~\bibnamefont {Landragin}}, \bibinfo {author}
  {\bibfnamefont {P.}~\bibnamefont {Bouyer}}, \bibinfo {author} {\bibfnamefont
  {A.}~\bibnamefont {Bertoldi}}, \bibinfo {author} {\bibfnamefont
  {R.}~\bibnamefont {Geiger}}, \ and\ \bibinfo {author} {\bibfnamefont
  {B.}~\bibnamefont {Canuel}},\ }\href {\doibase 10.1088/1361-6455/aa7592}
  {\bibfield  {journal} {\bibinfo  {journal} {Journal of Physics B: Atomic,
  Molecular and Optical Physics}\ }\textbf {\bibinfo {volume} {50}},\ \bibinfo
  {pages} {155002} (\bibinfo {year} {2017})}\BibitemShut {NoStop}%
\end{thebibliography}%
\bibliographystyle{aipnum4-1}
\label{biblio}


\onecolumngrid
\appendix
\newpage

\section*{SUPPLEMENTARY MATERIAL}

\section*{Intensity profile: additional data and discussion}

\begin{figure}[!h]
	\centering
	\includegraphics[width=\linewidth]{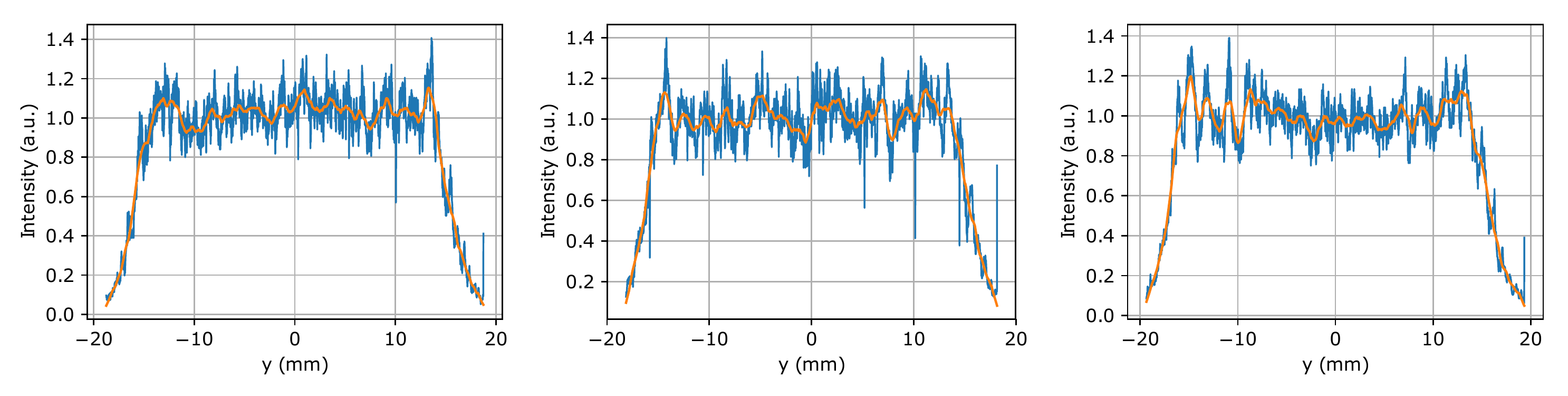}
	\caption{Cut of the intensity profile for various propagation distances. Left: 0~cm. Center: 40~cm (Fig.~1 d) of the main text). Right: 130~cm.}
	\label{fig:phase_maps}
\end{figure}

While we did not investigate in our work the origin of the residual intensity noise, it is probable that it originates from the roughness of the optics used in the beam shaper (containing 7 (aspheric) lenses) and in the input collimator and expander, which induces scattered light that interferes with the main beam. Improved surface quality of these optical surfaces may  result in a lower intensity noise. For comparison, a $5\%$ rms intensity noise was achieved in Ref.~\cite{Hoffnagle2000}, were only 2 aspheric lenses were used in the beam shaper.

\section*{Relative phase: additional data and discussion}
We show in Fig.~\ref{fig:phase_maps} the relative phase maps extracted from the measurement with the asymmetric Michelson interferometer for different distances of propagation (i.e. various  differences in arm length between the two arms of the interferometer).
The first column shows the measurement for zero propagation distance and allows to assess the limit of the method given by the quality of the optics. 

Numerical calculations show evidence that this relative phase measurement represents an upper bound on the relative phase inhomogeneities of the \tophat beam. When simulating an asymmetric Michelson interferometer injected with a perfect \tophat beam and with optics of surface flatness of $\lambda/10$ PV (reflecting  the value specified by the  manufacturer), we recover relative phase maps with inhomogeneities of typically $\lambda/3$ PV for a difference of propagation of $60$~cm, similar to the values reported in Fig.~\ref{fig:phase_maps}. Moreover, when numerically propagating a \tophat beam with an  intensity noise similar to the one shown in Fig.~1c) of the main text, we find relative phase inhomogeneities lower than $\lambda/16$ PV (and $\lambda/150$ rms) over 1~m of propagation.

\begin{figure}[!h]
	\centering
	\includegraphics[width=\linewidth]{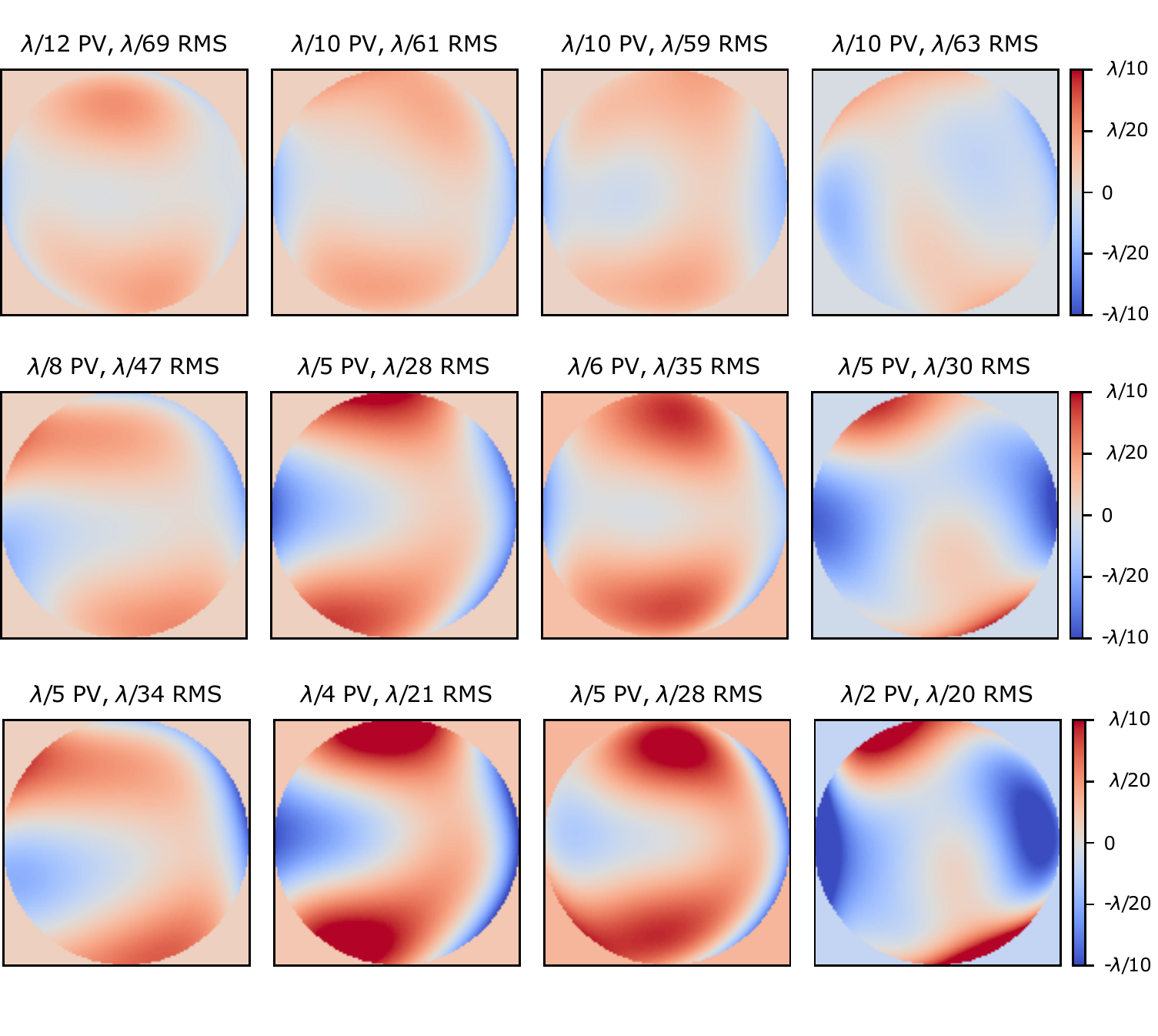}
	\caption{Relative phase maps for various distances of propagation and pupil diameters. From top to bottom, the pupil diameter is 15~mm, 20~mm and 28~mm. From left to right, the propagation distances are 0~cm, 30~cm, 70~cm (Fig.~1e) of the main text), and  110~cm.}
	\label{fig:phase_maps}
\end{figure}

Still, we computed the impact of such relative phase inhomogeneities on the bias of a 3-light-pulse atom interferometer with pulse separation time $T$. To this end, we numerically propagated an atom cloud with the velocity distribution used previously (see first section)  in the \tophat beam, and extracted the interferometer phase. For a pulse separation time as long as $T=0.5$~s, we find a bias of typically 40~mrad.  

\section*{Relative phase for an ideal \tophat beam}
\begin{figure}[!h]
	\centering
	\includegraphics[width=\linewidth]{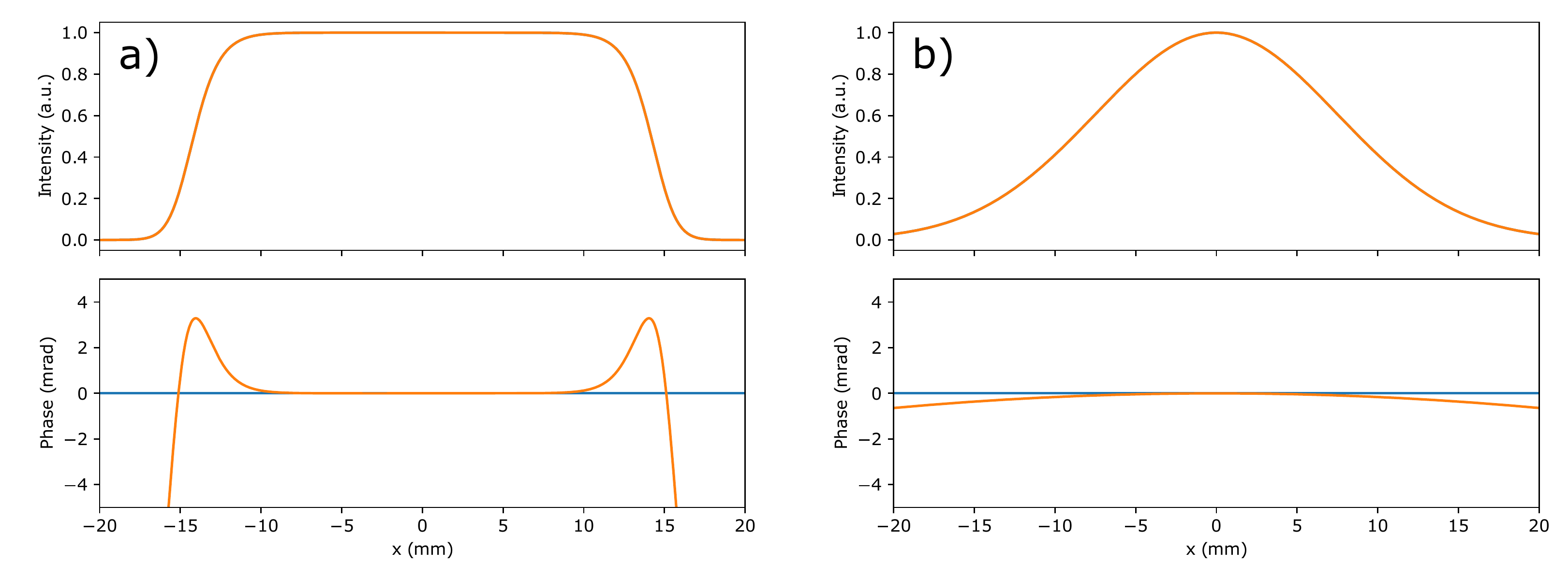}
	\caption{Relative phase of a propagated \tophat beam (a) compared to that of a Gaussian beam (b) for a distance of propagation of 30~cm. The top rows show the intensity profile and the bottom rows the relative phase. The beams have a circular symmetry and we show here the profiles along a line.}
	\label{fig:tophat_ideal_phase}
\end{figure}

As described in, e.g. Ref.~\cite{Gori1994}, a flattened Gaussian beam such as our \tophat beam can be expressed as a sum of Laguerre-Gauss modes, LG$_n$ (with mode index $n$), which propagation can be analytically computed. In Fig.~\ref{fig:tophat_ideal_phase}, we show as an illustration the relative phase of an ideal \tophat beam (of 30~mm FWHM) for a distance of propagation of 30~cm, and compare it to the relative phase of an ideal Gaussian beam of similar size (30~mm diameter at $1/e^2$). The relative phase inhomogeneities are well below 1~mrad until the edge of the \tophat beam is reached.

\section*{Model for the Rabi oscillations}
To model the Rabi oscillations, we employ a  Monte-Carlo simulation where we generate an ensemble of atoms with individual velocities following the   distribution measured with the Doppler-sensitive Raman transitions (corresponding to a 3D temperature of $1.2 \ \mu$K). The individual transition probabilities are calculated according to
\begin{equation}
	P(\vec{r}, v) = \frac{\Omega(\vec{r})^2}{\Omega(\vec{r})^2 + \delta_D^2} \sin^2 \left[ \frac{\tau}{2}\sqrt{\Omega(\vec{r})^2 + \delta_D^2}  \right],
\end{equation}
where $\Omega(\vec{r})/2\pi$ is the two-photon Rabi frequency, proportional to the local intensity of the beam at the position $\vec{r}$ of the atom, $\tau$ the pulse duration, and $\delta_D = k_{eff} v$ the two photon  detuning, with $k_{eff} = 4\pi / \lambda$ the two-photon wave-vector  and $v$ the velocity of the atom in the  direction of the beam.
Damping of the Rabi oscillations results from the average of many sinusoids with different Rabi frequencies and/or detunings.
The simulation accounts for the finite detection region of the atoms (modeled as a rectangle of 30~mm by 30~mm in the plane transverse to gravity), and for spontaneous emission.
The peak intensity is fitted on the upwards Rabi oscillation and fixed to this value for the downward oscillations. 
The model reproduces well the  data, and allows to assess the intensity homogeneity of the \tophat beam.

\section*{Gain in atom inteferometer performance with the \tophat beam: numerical examples}
We present here two examples of  performance improvement offered by the use of \tophat beams over Gaussian beams in different  atom interferometer geometries.\\
\\
 \textbf{Gain in contrast in a 4-pulse atom interferometer.}
The model for the Rabi oscillations presented in the previous section matches well the experimental data (Fig.~4 of the main text), and can thus be used to simulate the contrast of an atom interferometer in a particular geometry. We simulate a 4-pulse atom interferometer with a total interrogation time of 800~ms as in Ref.~\cite{Dutta2016}, using  the same velocity distribution as before (Cesium atoms, temperature 1.2~$\mu$K). 
We take into account the finite laser power which we keep constant between the \tophat and Gaussian beams.
A compromise has to be operated: while a large beam size allows to address more atoms (after expansion), it results in a decrease of  the peak intensity, which is detrimental to the transfer efficiency owing to the velocity selectivity of counter-propagating Raman transitions. Therefore, we calculate the optimal size of the beams that provides the highest contrast. We assume that the two beams have the same size and receive the same optical power. We initialize the optimization by considering a laser power corresponding to a peak two-photon Rabi frequency of 25~kHz for 20~mm waist Gaussian beams.
 After optimization, we find a maximum contrast of  $35\%$ for optimal Gaussian beams of 16~mm waist; the maximum contrast is  $56\%$ for optimal \tophat beams of 24~mm FWHM (the FWHM of the \tophat beam presented in the main text is 31.7~mm).\\ 
 \\
 \textbf{Large Momentum Transfer (LMT) atom-optics.} We numerically evaluated the impact of Rabi frequency inhomogeneities on the efficiency of LMT Bragg transition, using the numerical model developed in our earlier works (e.g. Ref.~\cite{riou_marginally_2017}). We find that, in the quasi-Bragg regime, a variation  of $10\%$ in the two-photon Rabi frequency (proportional to laser intensity) leads to $20\%$ of variation in the $\pi$ pulse efficiency for a $6\hbar k$ Bragg transition (as used, e.g., in Ref.~\cite{Mazzoni2015}). Maintaining a $10\%$ rms intensity variation within a Gaussian beam amounts to use only $25\%$ of the power corresponding to the central part.
In comparison, employing a \tophat beam with the same intensity profile  as the one reported in our work (Fig.~1d)) allows to use more than $75\%$ of the power.

\end{document}